\documentclass[a4paper,11pt]{article}
\usepackage{jhepmod}

\title{Collective Properties of Gravitational Perturbations in Thermal Environments on a Dynamical Background}

\author[1]{Atsuhisa Ota}
\affiliation[1]{Department of Physics and Chongqing Key Laboratory for Strongly Coupled Physics, \\
Chongqing University, Chongqing 401331, People's Republic of China}

\emailAdd{aota@cqu.edu.cn}

\abstract{
A thermal plasma dresses soft gauge fields, turning them into collective modes with a plasmon gap and Landau damping.
In this paper, we ask whether a thermal radiation plasma does the same to gravitational perturbations.
For gravity, the thermal soft scale coincides with the Hubble rate.
Any such gap therefore lies at the horizon scale, where the expansion of the background cannot be neglected.
We model radiation by a conformal scalar and compute its linear response to metric perturbations directly on a flat FLRW background.
We test this response against the diffeomorphism Ward identities.
The adiabatic dispersion relation yields a gapped collective mode with $\omega_p^2 = 8\mathcal H^2/5$, where $\mathcal H$ is the conformal Hubble rate.
The imaginary part of its frequency, however, is dominated by the background expansion and does not represent Landau damping.
In the radiation background, the tensor modes obey a real-time integro-differential equation that reproduces Weinberg's damping of gravitational waves.
The initial-time boundary of the retarded response keeps superhorizon modes frozen until horizon reentry despite the gap.
The superhorizon growth found in earlier analyses arises from discarding the memory of the retarded response and from a local-equilibrium state prescription that violates the Ward identities.
More generally, the Ward identities forbid any local state response beyond the collisionless kernel.
}

\begin{document}
\maketitle

\section{Introduction}
The propagation of waves depends on the properties of the surrounding medium.
Gravitational perturbations are no exception, but their coupling to matter is
suppressed by the Planck scale, so medium effects are ordinarily negligible.
The early universe, however, was filled with a hot, dense plasma close to thermal
equilibrium.
Its large energy density may offset this suppression.
The effect of such a medium on gravitational perturbations is therefore a
quantitative question.

At finite temperature, gauge theories possess a soft scale at which naive
perturbation theory breaks down~\cite{Braaten:1990az,Pisarski:1988vd,Taylor:1990ia}.
At this scale, the appropriate infrared degrees of freedom are dressed by the
medium and propagate as collective modes.
They develop a gap in the dispersion relation, the plasmon mass, and are damped through resonance with particles that
keep pace with the wave~\cite{Bellac:2011kqa}.
This gap is consistent with gauge symmetry, which forbids a mass term in
the Lagrangian but not a gap in the dispersion relation.

For gravity, the corresponding soft scale is $\kappa T^2$, where $\kappa$ is the
gravitational coupling and $T$ is the radiation
temperature~\cite{Bellac:2011kqa,Brandt:1993bj,Francisco:2016rtf}.
The Friedmann equation sets the Hubble rate at the same scale,
$H\sim \kappa T^2$.
The horizon scale therefore plays the role of the plasmon mass.
Taken literally, this would forbid superhorizon gravitational waves, whose frequency at generation during inflation lies below the gap, in conflict with standard cosmology, which predicts that superhorizon tensor modes exist and stay constant~\cite{Mukhanov:2005sc}.
Since the gap is set by the expansion rate itself, the question cannot be settled
on a static background.
It must be posed on a dynamical one.

\medskip
In this paper, we therefore compute the linear response of thermal radiation to
gravitational perturbations on a spatially flat Friedmann--Lema\^itre--Robertson--Walker (FLRW) background.
We work in the canonical operator formalism of quantum field theory and do not fix a gauge for the
metric perturbations.
A conformal scalar serves as a minimal proxy for radiation.
We retain both the nonlocal retarded response and the local contact response.
The contact response is required because the stress tensor depends explicitly on
the metric.
Discarding such contributions is known to violate gauge invariance in analogous
condensed-matter settings~\cite{PhysRevB.71.104511,PhysRevB.86.245309}.
Canonical time ordering produces Schwinger terms at intermediate
stages~\cite{Schwinger:1959xd,Deser:1967zzf}.
We show explicitly that the contact response cancels them.

A dynamical background is likewise essential for implementing the constraints of
diffeomorphism symmetry.
Analyses carried out on a static Minkowski background~\cite{Rebhan:1990yr,Frenkel:1990dw,Brandt:1991qn,Brandt:1993bj,Nachbagauer:1995wn} are insufficient for this
purpose because flat spacetime is not a solution of the Einstein equation with a non-vanishing stress tensor.
The form of the Ward identity depends on how the metric perturbations are defined.
We therefore construct the Ward identities on an FLRW background in each of the
common conventions.
Together with the Weyl identities of conformal matter, they provide a test of the
construction.

We treat the cosmic expansion as slow and solve the resulting adiabatic
dispersion relation.
The expected plasma features emerge.
The gravitational perturbation supports a gapped collective mode with
$\omega_p^2 = 8\mathcal H^2/5$, where $\mathcal H$ is the conformal Hubble rate.
The frequency of this mode also acquires an imaginary part, whose leading contribution reflects the expansion rather than the branch cut.
This imaginary part cannot be interpreted as a damping rate, and the collective pole cannot be resolved from the Landau cut within a Hubble time.
The cosmological problem, however, is a real-time one on a dynamical background.
It takes the form of an integro-differential equation.
In the hard-thermal-loop (HTL) limit, this equation agrees with the result obtained from kinetic
theory~\cite{Rebhan:1994zw,Weinberg:2003ur,Watanabe:2006qe}.
The damping in its solution is carried by the Landau cut rather than by the collective pole.
Once the infrared effective equation is derived with the initial-time boundary properly
retained, the boundary acts as a source for long-wavelength gravitational waves.
Superhorizon modes then remain frozen until horizon
reentry despite the gap~\cite{Weinberg:2003sw}.
Near the horizon, the correction to the tensor amplitude reaches a few tens of percent.
The superhorizon secular growth found in our earlier analyses~\cite{Ota:2023iyh,Ota:2024idm,Ota:2025yeu} has two origins.
The secular behavior results from discarding the memory of the retarded kernel, while the growth results from a state prescription that violates the Ward identities.
By retaining the memory of the retarded kernel and starting from global equilibrium, we remove both causes of the secular growth.
More generally, we show that the Ward identities forbid adding any local state response to the collisionless kernel at leading HTL order.

\medskip
The remainder of this paper is organized as follows.
In Sec.~\ref{seclinresqft}, we set up the linear-response framework for metric
perturbations on an FLRW background.
In Sec.~\ref{secdiff}, we introduce an effective action for the linear response,
derive the diffeomorphism Ward identities on an FLRW background, and explain
their dependence on quadratic field redefinitions.
In Sec.~\ref{symtest}, we formulate a combined diffeomorphism and Weyl symmetry
test and apply it to our linear-response construction and to the local Gibbs ensemble.
In Sec.~\ref{seceffdyn}, we apply this framework to tensor perturbations and
derive the integro-differential evolution equation and the adiabatic dispersion
relation.
We conclude in the final section.

\section{Linear response in quantum field theory}
\label{seclinresqft}

Let us consider the gravitational linear response of a radiation background driven by a conformal scalar field in thermal equilibrium.
In this section, we derive the leading HTL stress-tensor response in three steps.
First, we define the linear response in the canonical operator formalism and emphasize that, in addition to the nonlocal retarded response, one must also include the local contact response.
Next, we evaluate the retarded and contact contributions in canonical time ordering, where the Schwinger terms arise at intermediate stages.
Finally, we combine these contributions to obtain the total response.
To the best of our knowledge, the operator-formalism calculation in an FLRW background presented here is new.

\subsection{Setup}

We expand the metric around the FLRW background as
\begin{align}
	g^{\mu\nu} =a^{-2}\left( \eta^{\mu\nu} + \kappa h^{\mu\nu}\right) ,~\kappa = \sqrt{32\pi G},~h_{\mu\nu} =\eta_{\mu\rho}\eta_{\nu\sigma} h^{\rho\sigma}, \label{mycon}
\end{align}
and consider a conformal scalar $\chi$:
\begin{align}
	S[\chi;g] =-\frac{1}{2} \int d^4 x \sqrt{-g}\left( g^{\mu\nu}\nabla_\mu \chi \nabla_\nu \chi +\frac{1}{6}R\chi^2\right),
\end{align}
where $\nabla_\mu$ is the covariant derivative with respect to $g_{\mu\nu}$.
In this paper we do not attempt a systematic field-theoretic treatment for arbitrary spins; we use a conformal scalar as a minimal proxy and note that explicit gravitational-wave response analyses are available for photons and Weyl fermions in Refs.~\cite{Frob:2025sfq,Ota:2026yzp}.

The stress tensor is defined by
\begin{align}
	T_{\mu\nu} = - \frac{2}{\sqrt{-g}}\frac{\delta S}{\delta g^{\mu\nu}}.
\end{align}
We quantize $\chi$ and write the stress-tensor operator in the Heisenberg picture as
\begin{align}
	\hat T_{\mu\nu} = \partial_\mu \hat \chi \partial_\nu \hat \chi - \frac{1}{2} g_{\mu\nu} g^{\rho\sigma}\partial_\rho \hat \chi \partial_\sigma \hat \chi + \frac{1}{6}\left(g_{\mu\nu}g^{\rho \sigma}\nabla_\rho \nabla_\sigma - \nabla_\mu \nabla_\nu +G_{\mu\nu} \right)\hat \chi^2,\label{deftmunu}
\end{align}
where $G_{\mu\nu}$ is the Einstein tensor.

We study the interaction-picture evolution of the stress tensor in a radiation background sourced by the scalar itself.
We define the interaction-picture operators by
\begin{align}
	\hat T^I_{\mu\nu} = \lim_{\kappa \to 0}\hat T_{\mu\nu},\qquad \hat \chi^I = \lim_{\kappa\to 0}\hat \chi,\label{defintpictfield}
\end{align}
and consider the interaction between the metric perturbation and the scalar field:
\begin{align}
	\hat H_{\rm int}  = \frac{\kappa}{2}\int d^3 x\, a^2 h^{\mu\nu}\hat T^I_{\mu\nu}+\mathcal O(\kappa^2).\label{def:Hint}
\end{align}
We adopt the interaction-picture convention commonly used in cosmological perturbation theory, in which the perturbative time evolution is assigned to the operators while the reference state is kept fixed in time.
The leading evolution of $\hat T_{\mu\nu}$ is described by linear response theory~\cite{Kubo:1957mj}.
We write
\begin{align}
	\hat T_{\mu\nu} = \hat T^I_{\mu\nu} +  \hat T^{R}_{\mu\nu} +   \hat T^{C}_{\mu\nu},\label{deflineres}
\end{align}
with
\begin{align}
	\hat T^{R}_{\mu\nu} &= i \int dy^0\,  \Theta(x^0-y^0) \left[\hat H_{\rm int}(y), \hat T^I_{\mu\nu}(x) \right],\label{defresp}
	 \\
	  \hat T^{C}_{\mu\nu} &=\int d^4 y\, h^{\rho \sigma}(y) \left(\frac{\delta \hat T_{\mu\nu}(x)}{\delta h^{\rho\sigma}(y)}\right)_{\chi,\pi_\chi,\kappa=0} .\label{mycont}
\end{align}
Here $\Theta$ is the step function.
The commutator term gives the retarded response through the dynamical evolution of $\chi$ and its conjugate momentum $\pi_{\chi}$.
Although this term represents a genuinely nonlocal response, it contains both fast and slow components, characterized by the thermal and external momentum scales, respectively.
In Sec.~\ref{seceffdyn}, we will show that the fast component reduces to a local mass-like contribution canceled in the on-shell background, while the slow component reproduces the linear response predicted by traditional kinetic theory.

The third term is the contact response arising from the explicit metric dependence of the stress tensor.
This term is analogous to the partial time-derivative term in the Heisenberg equation, and the variation must be taken at fixed canonical variables.
The role of contact terms in linear response is familiar from other settings.
In condensed-matter applications, discarding contact contributions, for example in optical conductivity~\cite{PhysRevB.71.104511} or Hall viscosity~\cite{PhysRevB.86.245309}, violates gauge invariance and leads to inconsistencies.
In the present calculation, the contact term is necessary to cancel the non-covariant Schwinger terms that arise from the retarded response and to satisfy the diffeomorphism Ward identity, as discussed shortly.

\medskip
For a given quantum state, we evaluate the expectation value of \eqref{deflineres} and identify the effective stress tensor that appears on the right-hand side of the Einstein equation,
\begin{align}
	G_{\mu\nu} = \frac{\kappa^2}{4} \langle \hat T_{\mu\nu}\rangle.\label{einseq}
\end{align}
Although the notion of a thermal state in curved spacetime is subtle, at leading order in the perturbative expansion we define the state such that the unperturbed stress tensor takes the perfect-fluid form.
That is, with 
\begin{align}
	\bar T_{\mu\nu} = a^2 (\bar \rho +\bar P) \delta^0_\mu \delta^0_\nu + a^2 \bar P \eta_{\mu\nu},\label{defhomstten}
\end{align}
we match the quantum state such that
\begin{align}
	\left. \langle \hat T^I_{\mu\nu}\rangle \right|_{\kappa=0} = \bar T_{\mu\nu}.
\end{align} 
This matching implies that we take the conformal Gibbs distribution in the leading order of $\kappa$.
For the conformal scalar, $\bar P = \bar \rho/3$.
This fixes the state at $h^{\mu\nu}=0$ only.
We revisit the dependence of the state on the metric perturbation at the end of this section.

One may instead set $a=1$, calculate in the flat background, and restore the scale factor at the end.
We keep $a\neq1$, because the Weyl rescaling generates a time boundary term in Eq.~\eqref{defresp} and changes which canonical variables are held fixed in Eq.~\eqref{mycont}.
This subtlety is implicit in the standard imaginary time calculation.

At one loop, the two-point function of $h^{\mu\nu}$ receives contributions from two diagrams.
The nonlocal diagram contains two three-point vertices joined by a radiation loop.
The local diagram contains a single four-point vertex from the $\mathcal O(\kappa^2)$ term in Eq.~\eqref{def:Hint}.
In the present approach, we compute the effective vertex rather than the correlation function.
The nonlocal diagram then corresponds to $\hat T^R_{\mu\nu}$, and the local diagram to $\hat T^C_{\mu\nu}$.
The two organizations differ in their counting.
The loop order of the correlation function does not coincide with the HTL expansion parameter.
Near the horizon, the thermal self-energy of order $\mathcal H^2$ is comparable to the kinetic term.
Retaining the leading HTL contribution in the correlation function then requires resumming the chain of one-loop self-energy insertions~\cite{Ota:2023iyh}.
The effective vertex obtained here incorporates this resummation at leading HTL order, so the resulting equation of motion is consistent in the leading HTL expansion.

\subsection{Retarded response}

Now, let us compute the retarded response in Eq.~\eqref{deflineres}.
From Eqs.~\eqref{def:Hint} and \eqref{defresp}, one finds
\begin{align}
	\langle \hat T^{R}_{\mu\nu} \rangle
	= \frac{i\kappa}{2}\int d^4y\, a(y^0)^2 \Theta(x^0-y^0)\,
	\langle [\hat T^I_{\rho \sigma}(y),\hat T^I_{\mu\nu}(x)] \rangle\, h^{\rho \sigma}(y). \label{linresp}
\end{align}
In the current form, the linear response is expressed as the convolution integral for the retarded kernel and the gravitational perturbations.
The convolution form makes causality manifest, but obscures the operator structure needed for the subsequent symmetry analysis. 
We therefore integrate by parts so that the remaining soft derivatives act directly on the metric perturbation. The resulting operator retains the nonlocality through inverse directional derivatives.
This form allows us to organize the bulk response in terms of a single-copy effective action and to test it against the Ward identities in the next section.

\medskip
We first simplify the response kernel~\eqref{linresp} in terms of the Pauli-Jordan function and the Keldysh Green function.
For the conformal scalar, it is sufficient to evaluate
\begin{align}
	X_{\mu\nu\rho\sigma}(x,y)
	\equiv \frac{i\kappa}{2} a(y^0)^2\Theta(x^0-y^0)
	\left\langle [\partial_\rho \hat \chi^I(y) \partial_\sigma \hat \chi^I(y) , \partial_\mu \hat \chi^I(x) \partial_\nu \hat \chi^I(x) ]\right \rangle.
\end{align}
In terms of $X_{\mu\nu\rho\sigma}$, the response~\eqref{linresp} is the trace-subtracted combination
\begin{align}
\left(X_{\mu\nu\rho\sigma}
	-\frac{1}{2}\eta_{\rho\sigma} \eta^{\alpha \beta}X_{\mu\nu \alpha\beta}
	-\frac{1}{2}\eta_{\mu\nu} \eta^{\alpha \beta}X_{\alpha\beta \rho \sigma}
	+ \frac{1}{4} \eta_{\mu\nu} \eta_{\rho\sigma} \eta^{\alpha \beta}\eta^{\gamma \delta}X_{\alpha\beta \gamma \delta}\right)\star h^{\rho \sigma},\label{defpp}
\end{align}
where $\star$ implies the convolution 
\begin{align}
	X_{\mu\nu\rho\sigma}\star h^{\rho \sigma} \equiv \int d^4 y X_{\mu\nu\rho\sigma}(x,y) h^{\rho \sigma}(y).
\end{align}
The improvement part of the conformal stress tensor carries additional derivative corrections.
These derivatives act on $\hat \chi^2$ rather than on a single $\hat \chi$, so they do not produce hard momenta and are subleading in the HTL expansion.
It is therefore sufficient to evaluate the $X_{\mu\nu\rho\sigma}$ and find the retarded response by using Eq.~\eqref{defpp} at the end.
Hereafter we always drop the subleading HTL correction unless otherwise stated.

With the Keldysh Green function and the Pauli-Jordan function defined as
\begin{align}
	G^K(x,y) &\equiv \langle \hat \chi^I(x)\hat \chi^I(y)\rangle+  \langle \hat \chi^I(y)\hat \chi^I(x)\rangle,\label{defgk}
\\
	G(x,y) &\equiv i[\hat \chi^I(x),\hat \chi^I(y)],
\end{align}
one finds
\begin{align}
	X_{\mu\nu\rho\sigma}
	= -\kappa a(y^0)^2 \Theta(x^0-y^0)\,
	\partial_{\langle \mu} \partial_\rho G(x,y)\, \partial_\nu \partial_{\sigma\rangle} G^K(x,y).
\end{align}
Here all partial derivatives are with respect to $y$ unless otherwise stated.  
We introduced the cyclic symmetrization
\begin{align}
	A_{\langle \mu\rho\nu\sigma \rangle}
	= \frac{1}{4}\left( A_{\mu\rho\nu\sigma} + A_{\rho\nu\sigma\mu} + A_{\nu\sigma\mu\rho}+ A_{\sigma\mu\rho\nu} \right).
\end{align}

\medskip
The next step is to express the Pauli-Jordan function in terms of the retarded Green function
\begin{align}
	G^R(x,y) \equiv \Theta(x^0-y^0) G(x,y).
\end{align}
Because we consider the retarded response of composite operators involving time derivatives, the non-commutativity of time differentiation with the step function produces Schwinger terms~\cite{Schwinger:1959xd,Deser:1967zzf}.
This is a consequence of canonical time ordering, and we will show that the contact contribution cancels the non-covariant terms.
To identify the Schwinger terms explicitly, we write
\begin{align}
	\int d^4 y\, \Theta(x^0-y^0)\partial_{\mu} \partial_\rho G(x,y)\cdots
	&=\int d^4 y\, \partial_{\mu} \partial_\rho [\Theta(x^0-y^0) G(x,y)]\cdots
	\notag \\
	&-\int d^4 y\, \partial_{\mu} \Theta(x^0-y^0)\, \partial_\rho  G(x,y)\cdots
	\notag \\
	&-\int d^4 y\, \partial_{\rho} \Theta(x^0-y^0)\, \partial_\mu  G(x,y)\cdots
	\notag \\
	&-\int d^4 y\, [\partial_{\rho} \partial_\mu \Theta(x^0-y^0) ]\, G(x,y)\cdots.  \label{shch}
\end{align}
The Pauli-Jordan function satisfies
\begin{align}
	\lim_{y^0\to x^0}\partial_{x^0}G(x,y) &=\frac{1}{a(x)^2} \delta^{(3)}(\vec x - \vec y),
	\\
	\lim_{y^0\to x^0} G(x,y) &= 0,
\end{align}
which simplify the Schwinger terms and give
\begin{align}
	X_{\mu\nu\rho\sigma}\star h^{\rho \sigma}
	&=\kappa \lim_{y\to x}\delta^0_{\langle \mu}\delta^0_{\rho}\partial_\nu \partial_{\sigma\rangle} G^K(x,y)h^{\rho \sigma}(y)
	\notag 
	\\
	&-\kappa \int d^4 y\, a(y^0)^2 \partial_{\langle \mu} \partial_\rho G^R(x,y)\, \partial_\nu \partial_{\sigma\rangle} G^K(x,y)h^{\rho \sigma}(y).\label{Pidd}
\end{align}

\medskip
In the FLRW background the Keldysh Green function is not invariant under time translation, while that of the rescaled scalar $\tilde \chi = a\chi$ is invariant because we take the conformal Gibbs distribution.
We have
\begin{align}
	G^K(x,y) = \frac{\tilde G^K(x,y)}{a(x^0)a(y^0)},
\end{align}
and the rescaled Keldysh Green function is written as
\begin{align}
	\tilde G^K (x,y)&= \int \frac{d^4 p}{(2\pi)^4} e^{i p \cdot  (x - y)}\tilde G^K(p),\label{GKft}
\end{align}
where the dot product is defined with respect to $\eta^{\mu\nu}$.
The same rescaling applies to $G^R$.
Then, the nonlocal term in Eq.~\eqref{Pidd} becomes
\begin{align}
	-\kappa \int d^4 y\,a(y^0)^2 \partial_{\langle \mu} \partial_\rho \left[\frac{ \tilde G^R(x,y)}{a(x^0)a(y^0)}\right]\, \partial_\nu \partial_{\sigma\rangle} \left[ \int \frac{d^4 p}{(2\pi)^4} e^{i p \cdot  (x - y)}\frac{\tilde G^K(p)}{a(x^0)a(y^0)} \right] h^{\rho \sigma}(y).
\end{align}
By integrating by parts in $y$, we find
\begin{align}
	  \kappa \frac{1}{a(x^0)^2}\int \frac{d^4 p}{(2\pi)^4}  \tilde G^K(p) \int d^4 y\,  \tilde G^R(x,y)  e^{i p \cdot  (x - y)} D_{\langle \mu} D_\rho \, p_\nu p_{\sigma\rangle}h^{\rho \sigma}(y),\label{unnpi}
\end{align}
where we introduced
\begin{align}
	D_\mu = \partial_\mu - ip_\mu .
\end{align}
Note that a derivative acting on the scale factor gives the Hubble parameter, which is suppressed as $\mathcal H/p \sim H/T$ in the HTL expansion.
The HTL expansion drops the soft scales relative to $T$, both $k/T$ and $\mathcal H/T$.
Then, we use the position-space resolvent identity
\begin{align}
	\int d^4 y \tilde G^R(x,y) e^{ip\cdot (x-y)} F(y) = - \frac{1}{[D \cdot D]^R} F(x),\label{iduseful}
\end{align}
with the superscript $R$ implying the retarded condition. 
One can check this identity by multiplying $[D \cdot D]^R$ and the left-hand side.
Using Eq.~\eqref{iduseful} in Eq.~\eqref{unnpi}, we find
\begin{align}
	-\kappa \frac{1}{a(x^0)^2}\int \frac{d^4 p}{(2\pi)^4} \tilde G^K(p) \frac{D_{\langle \mu} D_\rho \, p_\nu p_{\sigma\rangle}}{[D \cdot D]^R}  h^{\rho\sigma}.
\end{align}
For on-shell hard momentum, $p\cdot p=0$, and in the HTL expansion we have
\begin{align}
	\frac{1}{D\cdot D }
	\approx -\frac{1}{2} \frac{1}{ip\cdot \partial}\left(1 +\frac{1}{2} \frac{\partial\cdot \partial}{ip\cdot \partial }  \right) .
\end{align}
Combining these results, the nonlocal term~\eqref{Pidd} can be recast as
\begin{align}
	\frac{\kappa}{4}\frac{1}{a(x^0)^2} \int \frac{d^4 p}{(2\pi)^4}  \tilde G^K(p) \left(2i \frac{p_\mu p_\rho  p_\sigma p_\nu}{[p\cdot \partial]^R}+  \frac{\partial^2 p_\mu p_\rho  p_\sigma p_\nu}{([p\cdot \partial]^R)^2} - 4 \frac{\partial_{\langle \mu} p_\rho  p_\sigma p_{\nu\rangle}}{[p\cdot \partial]^R}\right) h^{\rho \sigma}.\label{eq241}
\end{align}
Note that $p\cdot \partial = \eta^{\mu\nu}p_\mu \partial_\nu$.
The first term is super-leading in the HTL.
Under $p_\mu\to-p_\mu$, $\tilde G^K(p)$ is even by Eq.~\eqref{defgk}, while the retarded inverse $1/[p\cdot\partial]^R$ is odd; the retarded boundary condition is unaffected by this transformation. Hence, the super-leading contribution vanishes.

To summarize, we obtain
\begin{align}
	X_{\mu\nu\rho\sigma}\star h^{\rho \sigma}
	=-\frac{\kappa}{2}a^2 \bar \Pi^R_{\mu\nu\rho\sigma}h^{\rho \sigma}-2\kappa\delta^0_{\langle \mu}\delta^0_{\rho}\bar T_{\nu\sigma\rangle}h^{\rho \sigma},\label{Pidd2}
\end{align}
where we defined~\cite{Rebhan:1990yr,Brandt:1991qn}
\begin{align}
	\bar \Pi^R_{\mu\nu\rho\sigma}
	&\equiv \frac{1}{2a^4}\int \frac{d^4 p}{(2\pi)^4}  \tilde G^K(p) \left(4 \frac{\partial_{\langle \mu} p_\rho  p_\sigma p_{\nu\rangle}}{[p\cdot \partial]^R}
	- \frac{\partial^2 p_\mu p_\rho  p_\sigma p_\nu}{([p\cdot \partial]^R)^2} \right),\label{defpibar}
	\\
	\bar T_{\mu\nu}&\equiv  \frac{1}{2a^2}\int \frac{d^4 p}{(2\pi)^4} p_\mu p_\nu \tilde G^K(p).
\end{align}
The leading perfect fluid form stress tensor~\eqref{defhomstten} is matched through Eq.~\eqref{GKft}.
For the Schwinger term we used
\begin{align}
	 \lim_{y\to x}\delta^0_{\langle \mu}\delta^0_{\rho}\partial_\nu \partial_{\sigma\rangle} G^K(x,y) = -2 \delta^0_{\langle \mu}\delta^0_{\rho}\bar T_{\nu \sigma\rangle}.\label{doneshwinger}
\end{align}
Finally, using Eqs.~\eqref{defpp} and \eqref{Pidd2}, one finds the linear response
\begin{align}
	\langle \hat T^{R}_{\mu\nu} \rangle=&- \frac{\kappa}{2} a^2 \bar \Pi^R_{\mu\nu\rho\sigma}h^{\rho \sigma}
	+\frac{\kappa}{2}\eta_{\mu\nu}\bar T_{\rho\sigma}h^{\rho \sigma}
	+\frac{\kappa}{2}h \bar T_{\mu\nu}
	\notag \\
	&
	 - \frac{\kappa a^2\bar \rho}{3}\delta^0_\mu \eta_{i\nu} h^{i0}-\frac{\kappa a^2\bar \rho}{3}\delta^0_\nu \eta_{i\mu} h^{i0}
	- \kappa \left( \delta^0_\mu \delta^0_\nu + \frac{1}{2}\eta_{\mu\nu}\right)a^2\bar \rho (h + 2 h^{00})
,     \label{234}
\end{align}
where $h = \eta_{\mu\nu}h^{\mu\nu}$.
The second line corresponds to the non-covariant Schwinger terms, which are not manifest in the imaginary-time formalism in the literature.
Note that we have dropped all HTL subleading corrections.

\subsection{Contact response}

We now evaluate the contact response $\hat T^C_{\mu\nu}$.
Define
\begin{align}
	\hat J_{\mu\nu} &\equiv \partial_\mu \hat \chi \partial_\nu \hat \chi,
	\\
	\hat J &\equiv  g^{\mu\nu} \hat J_{\mu\nu}.
\end{align}
Then the stress-tensor operator can be written as
\begin{align}
	\hat T_{\mu\nu} = \hat J_{\mu\nu} - \frac{1}{2} g_{\mu\nu} \hat J.
\end{align}
Since we work in the canonical formalism, the variation of the stress tensor is taken at fixed $\hat \chi$ and its conjugate momentum $\hat \pi_\chi$.
The conjugate momentum is
\begin{align}
	\hat \pi_\chi = -\sqrt{-g}\, g^{0\mu} \partial_\mu \hat \chi.
\end{align}
Holding $\hat \pi_\chi$ fixed makes $\partial_0 \hat \chi$ depend on the metric, while $\partial_i \hat \chi$ does not.
Imposing $\delta \hat \pi_\chi =0$ and $\delta \hat \chi=0$ gives
\begin{align}
	\delta \hat J_{ij} &=0,
	\\
	\delta \hat J_{i0} &=\kappa \delta h^{0j}\hat J_{ji} + \frac{\kappa}{2} \left( \delta h^{00} + \delta_{jk} \delta h^{jk} \right)  \hat J_{i0},
	\\
	\delta \hat J_{00} & = 2 \kappa \delta h^{0i}\hat J_{i0} +  \kappa \left( \delta h^{00} +\delta_{jk} \delta h^{jk} \right)  \hat J_{00},
	\\
	\delta \hat J &=a^{-2}\kappa\, \delta h^{\mu\nu}\hat J_{\mu\nu}-a^{-2}\delta \hat J_{00}.
\end{align}
The variation of the metric perturbation is
\begin{align}
	\frac{\delta h_{\mu\nu}(y)}{\delta h^{\rho \sigma}(x)}
	&=\delta^{(4)}(x-y) \frac{1}{2}(\eta_{\mu\rho}\eta_{\nu\sigma} + \eta_{\nu\rho}\eta_{\mu\sigma}),
\\
	\frac{\delta h^{\alpha \beta}(y)}{\delta h^{\rho \sigma}(x)}
	&= \delta^{(4)}(x-y)\frac{1}{2} \left(\delta^\alpha_{\rho} \delta^\beta_{\sigma} + \delta^\alpha_{\sigma} \delta^\beta_{\rho}\right).
\end{align}
Using these relations, we obtain
\begin{align}
	\langle \hat T^C_{\mu\nu}\rangle
	=- \frac{\kappa}{2}\eta_{\mu\nu}\bar T_{\rho \sigma} h^{\rho \sigma}
			 +\frac{\kappa a^2\bar \rho}{3} \delta^0_\mu \eta_{i\nu} h^{i0}
			 +\frac{\kappa a^2\bar \rho}{3} \delta^0_\nu \eta_{i\mu} h^{i0}
			 +\kappa \left( \delta^0_\mu \delta^0_\nu  +\frac{1}{2}\eta_{\mu\nu}\right) a^2\bar \rho (h + 2 h^{00})
	,\label{conteq}
\end{align}
which cancels the non-covariant Schwinger terms in the retarded response.

Schwinger terms appear because the canonical formalism employs the time ordering $T$.
If instead one evaluates the retarded response in the Lagrangian formalism via the path integral, the covariant time ordering $T^*$ is selected automatically.
In that case, the non-covariant Schwinger terms do not arise in $\hat T^{R}_{\mu\nu}$.
The variation defining the contact term is then taken at fixed $\hat J_{\mu\nu}$, so that only the explicit metric factors are varied.
We have explicitly checked that the $T$ and $T^*$ prescriptions agree in the present case.

\subsection{Summary: the total response}

Let us summarize the calculation presented in this section.
Starting from the linear response~\eqref{deflineres}, we evaluated the retarded response~\eqref{defresp} and the contact response~\eqref{mycont} in the unperturbed conformal Gibbs state.
The retarded and contact contributions are given by Eqs.~\eqref{234} and \eqref{conteq}, respectively.
Substituting Eqs.~\eqref{234} and~\eqref{conteq} into Eq.~\eqref{deflineres}, we obtain the effective stress tensor:
\begin{align}
	\langle \hat T_{\mu\nu}\rangle
	= \left(1 + \frac{\kappa}{2}h \right)\bar T_{\mu\nu}
	- \frac{\kappa}{2}a^2 (\bar \Pi^R_{\mu\nu\rho\sigma} + \bar \Pi^S_{\mu\nu\rho\sigma}) h^{\rho \sigma} .\label{defefferesukt}
\end{align}
The first term is separated from the self-energy since it represents the geometric response due to the volume perturbations.
In the intermediate steps, Schwinger terms appear as a consequence of canonical time ordering, but they cancel in the final result.
Fluctuations in the quantum state are added as the residue $\bar{\Pi}^{S}_{\mu\nu\rho\sigma}$, which is not zero in general.
An instantaneous distribution, such as the local Gibbs distribution, gives rise to a nonzero bulk contribution in $\bar \Pi^S_{\mu\nu\rho\sigma}$, and this bulk contribution shifts the local term in the response.
The boundary data specifying the initial conditions appear as a source in the equation of motion, but they do not modify the kernel as we discuss below.
The nonlocal part of our retarded response agrees with the kernel obtained by the standard retarded analytic continuation of the imaginary-time result~\cite{Rebhan:1990yr,Brandt:1991qn}.
In the present derivation, no analytic continuation is required because causality is built in from the outset.
The contact diagram in the path-integral formalism corresponds to the contact term defined at fixed $\hat J_{\mu\nu}$, and the total response agrees between the two formalisms.
In the next section, we discuss how to construct the effective action for the linear response to study the diffeomorphism invariance of our result.
Readers who are interested in solving the effective Einstein equation~\eqref{einseq} may jump to Sec.~\ref{seceffdyn}.

\section{Effective action and Ward identities}
\label{secdiff}

In this section, we discuss the symmetry structure of the linear response we obtained.
We examine whether our response~\eqref{defefferesukt} is consistent with diffeomorphism invariance.
For the symmetry argument, normally it is convenient to find the effective action and derive the Ward identities.
In what follows, we construct the effective action for the bulk response and discuss the constraints on it.

\subsection{Effective action for the bulk kernel}

Let us introduce a generating functional of the stress tensor through 
\begin{align}
	\langle T_{\mu\nu} \rangle =-\frac{2}{\sqrt{-g}}\frac{\delta \Gamma^R}{\delta g^{\mu\nu}}
	.\label{defEMT}
\end{align}
Then, one can formally integrate the effective stress tensor and find
\begin{align}
	\Gamma^R = \int d^4 x\,\left[-\frac{\kappa}{2} a^2 \bar T_{\mu\nu} h^{\mu\nu}
	+ \frac{\kappa^2}{8} h^{\mu\nu} a^4 (\bar \Pi^R_{\mu\nu\rho\sigma} + \bar \Pi^S_{\mu\nu\rho\sigma}) h^{\rho\sigma} \right].\label{eqdefGamma}
\end{align}
Varying the quadratic form in Eq.~\eqref{eqdefGamma} reproduces Eq.~\eqref{defefferesukt} through Eq.~\eqref{defEMT} when $a^4(\bar\Pi^R+\bar\Pi^S)$ is symmetric under the integration by parts in $x$.
The second variation of an action has this symmetry, whereas a retarded kernel generally does not.
The matter is collisionless, and damping through phase mixing does not imply microscopic collisional dissipation.
The asymmetry of the kernel thus arises only from the retarded condition and the Landau contour.
Without these prescriptions, the bulk kernel is formally self-adjoint under integration by parts and defines a single-copy action, which we denote by $\Gamma$.
We also require the bulk state response $a^4\bar\Pi^S$ to have this symmetry.

With a local gravity action $S_g$, the total action is then
\begin{align}
	S_{\rm tot} = S_g + \Gamma.\label{stot}
\end{align}
Gravitational perturbations are now dynamical.
After varying $S_{\rm tot}$, we impose the retarded condition on the Green function and take the Landau contour for the phase space integral; the causal response and any Landau damping associated with phase mixing are determined at this stage.
The mean response can be obtained in this way without introducing doubled fields.
To retain the causal prescription and initial data within the variational formulation itself, one uses doubled fields on the closed time path.
The matching prescription for the Schwinger-Keldysh effective action $\Gamma^{\rm SK}$ is
\begin{align}
	\langle T_{\mu\nu} \rangle =-\frac{2}{\sqrt{-g}}\left.\frac{\delta \Gamma^{\rm SK}}{\delta g_A^{\mu\nu}}\right|_{g_A=0}.\label{defEMTsk}
\end{align}
Here $g_A^{\mu\nu}=g_+^{\mu\nu}-g_-^{\mu\nu}$ is the difference between the inverse metrics on the two branches, and $g^{\mu\nu}=(g_+^{\mu\nu}+g_-^{\mu\nu})/2$.
The classical limit $g_A=0$ is taken after the variation.
With $g_A^{\mu\nu} =a^{-2} \kappa h_A^{\mu\nu}$, we obtain
\begin{align}
	\Gamma^{\rm SK} = \int d^4 x\, \left[-\frac{\kappa}{2}h_A^{\mu\nu} a^2\bar T_{\mu\nu} 
	+ \frac{\kappa^2}{4} h_A^{\mu\nu} a^4 (\bar \Pi^R_{\mu\nu\rho\sigma} + \bar \Pi^S_{\mu\nu\rho\sigma}) h^{\rho\sigma} + \mathcal O(h_A^2,h_Ah^2)\right].\label{eqdefGammaSK}
\end{align}
The omitted terms quadratic in $h_A$ contain the noise kernel, which is not determined here and does not contribute to the classical response at $h_A=0$.
Terms of order $h_Ah^2$ describe nonlinear response.
The symmetry structure there is richer, since the generators are doubled, and the initial condition can itself be tested.
We do not need it here, and we work with the single-copy action for the bulk kernel.

\subsection{Diffeomorphism Ward identity}

The background geometry plays an essential role in the thermal response to gravitational perturbations, yet much of the literature works in the thermal Minkowski limit~\cite{Rebhan:1990yr,Frenkel:1990dw,Brandt:1993bj,Nachbagauer:1995wn}. 
Although the general form of the Ward identities on an FLRW background is known~\cite{Tan:2021ibs}, their implications have not been made explicit within the parametrization~\eqref{mycon}: expanded about FLRW, the identity does not reduce to a transversality condition for the self-energy alone, but mixes it with the tadpole.
We further show that violations of the Weyl identity feed into the FLRW-specific term in the diffeomorphism Ward identity.
We derive the resulting identities explicitly in this section.

\medskip
Under the diffeomorphism $x^\mu \to x^\mu - \kappa \xi^\mu$,
\begin{align}
	g^{\mu\nu} \to g^{\mu\nu} + \kappa (\nabla^\mu \xi^\nu + \nabla^\nu \xi^\mu) + \mathcal O(\kappa^2).
\end{align}
The effective action $\Gamma$ varies as~\cite{Rebhan:1990yr}
\begin{align}
	\delta_\xi \Gamma = 2\kappa\int d^4 x\, \frac{ \delta \Gamma }{\delta g^{\mu\nu}}\nabla^{\mu} \xi^{\nu} .\label{eqvarxi}
\end{align}
Here $\xi^\mu$ is an arbitrary infinitesimal vector field that vanishes on the boundary.
If $\Gamma$ is diffeomorphism invariant, then $\delta_\xi \Gamma=0$ up to total derivatives, which implies local energy-momentum conservation.
We now expand Eq.~\eqref{eqvarxi} around a flat FLRW background and define
\begin{align}
	\bar T_{\mu\nu} & = - \frac{2}{\sqrt{-g}}\frac{ \delta \Gamma }{\delta g^{\mu\nu}(x)} \Bigg |_{\kappa=0},\label{eqstress}
	\\
	\frac{a^2(x)}{a^2(y)}\left[\bar \Pi_{\mu\nu\rho\sigma}\delta^{(4)}(x-y)\right] &=
	\frac{2}{\sqrt{-g(x)}}\frac{2}{\sqrt{-g(y)}}\frac{ \delta^2 \Gamma }{\delta g^{\rho \sigma}(y)\delta g^{\mu\nu}(x)} \Bigg |_{\kappa=0}.\label{pidefdiag}
\end{align}
The quantities $\bar T_{\mu\nu}$ and $\bar \Pi_{\mu\nu\rho\sigma}$ agree with the stress tensor and self-energy defined in the previous section after dropping the retarded condition and the initial condition.
A nonlocal self-energy cannot in general be written with a delta function as in Eq.~\eqref{pidefdiag}.
The kernel in Eq.~\eqref{defpibar} is constructed from derivatives and their retarded inverses, so it admits this representation as an operator acting on the $x$ dependence of $\delta^{(4)}(x-y)$.
The operator does not act on the scale factors outside the square brackets.
 
Expanding the covariant derivative in Eq.~\eqref{eqvarxi}, we find
\begin{align}
 \delta_\xi\Gamma
 =2\int d^4x\,\frac{\delta\Gamma}{\delta h^{\mu\nu}}
 \Big[&\partial^\mu\xi^\nu+\eta^{\mu\nu}\xi^\alpha\partial_\alpha\ln a\Big]
 +\mathcal O(\kappa^2).
 \label{wi_variation_expanded}
\end{align}
At leading order in $\kappa$, for an arbitrary test function $\xi^\mu$, integration by parts yields
\begin{align}
	\partial^\mu ( a^2 \bar T_{\mu\nu} ) - a^2 \eta^{\rho\sigma}\bar T_{\rho\sigma} \partial_\nu \ln a = 0.\label{WI1}
\end{align}
These are the continuity and Euler equations for the background matter in an expanding background.

From Eqs.~\eqref{eqstress} and \eqref{pidefdiag}, we have 
\begin{align}
	\frac{\delta\Gamma}{\delta h^{\mu\nu}}
=
-\frac{\kappa}{2}a^2\bar T_{\mu\nu}
+\frac{\kappa^2}{4}a^4
\bar\Pi_{\mu\nu\rho\sigma}h^{\rho\sigma}
+\mathcal O(\kappa^3).
\end{align}
When Eq.~\eqref{WI1} is imposed, the next-to-leading order identity reduces to
\begin{align}
 0&=2\kappa^2\int d^4x\,
 \left(
 \frac{1}{4}a^4
\bar\Pi_{\mu\nu\rho\sigma}h^{\rho\sigma}
 \right)
 \Big[\partial^\mu\xi^\nu+\eta^{\mu\nu}\xi^\alpha\partial_\alpha\ln a\Big]
 \notag \\
 &+2\kappa^2\int d^4x\,
 \left(
-\frac{1}{2}a^2\bar T_{\mu\nu} 
 \right)
 \Big[h^{\mu\alpha}\partial_\alpha\xi^\nu
 -\frac{1}{2}\xi^\alpha\partial_\alpha h^{\mu\nu}
 + h^{\mu\nu}\xi^\alpha\partial_\alpha\ln a\Big]
 +\mathcal O(\kappa^3).
 \label{wi_variation_expanded_on_BG}
\end{align}
This identity must hold for arbitrary $\xi^\mu$ and $h^{\rho\sigma}$.
Hence the operator identity acting on $h^{\rho \sigma}$ is recast as
\begin{align}
 &\partial^\mu \circ (a^4\bar \Pi_{\mu\nu\rho\sigma})
 - \partial_\rho \circ \left( a^2 \bar T_{\sigma\nu}\right)
 - \partial_\sigma \circ \left( a^2 \bar T_{\rho \nu}\right)
 - a^2 \bar T_{\sigma\rho} \partial_{\nu} 
 \notag 
 \\
 &
 + 2 a^2 (\partial_\nu \ln a) \left( \bar T_{\rho \sigma}  -  \frac{1}{2} a^2 \eta^{\alpha \beta}\bar \Pi_{\alpha \beta \rho\sigma} \right) =0,\label{mywi}
\end{align}
where we defined 
\begin{align}
	\partial^\mu \circ (a^4\bar \Pi_{\mu\nu\rho\sigma}) h^{\rho \sigma} &\equiv  \partial^\mu (a^4\bar \Pi_{\mu\nu\rho\sigma}h^{\rho \sigma} ),
	\\
	\partial_\rho \circ \left( a^2 \bar T_{\sigma\nu}\right) h^{\rho \sigma} & \equiv \partial_\rho \left( a^2 \bar T_{\sigma\nu}h^{\rho \sigma} \right).
\end{align}
The background stress tensor $\bar T_{\mu\nu}$ is treated as a c-number.
Because the stress tensor is nonzero, the identity is not a simple transversality condition with respect to partial derivatives.
The last term is specific to an expanding background: $\partial_\mu \ln a$ is the conformal Hubble parameter and does not vanish on FLRW.
In fact, for a conformal field the Weyl identities
\begin{align}
	\eta^{\mu \nu} \bar T_{\mu\nu} &=0,\label{wel1}
	\\
	\bar T_{\mu\nu}  -\frac{1}{2}  a^2 \eta^{\rho\sigma} \bar \Pi_{\rho\sigma \mu\nu} & = 0\label{wel2}
\end{align}
are satisfied.
The FLRW-specific term therefore vanishes.
A violation of Eq.~\eqref{wel2} would appear directly in this term.

\subsection{Field-redefinition dependence}\label{subsec:fieldredef}
The perturbative expansion of $\Gamma$ is not unique.
We now give the identities in the two conventions used in the literature.
One may define the stress tensor and self-energy as~\cite{Rebhan:1990yr}
\begin{align}
	\tilde T^{\mu\nu} & = \frac{2}{\sqrt{-g}}\frac{ \delta \Gamma }{\delta g_{\mu\nu}(x)} \Bigg |_{\kappa=0},\label{unpi}
	\\
	\frac{a^6(x)}{a^6(y)}\left[\tilde \Pi^{\mu\nu\rho\sigma}\delta^{(4)}(x-y)\right] &=
	\frac{2}{\sqrt{-g(x)}}\frac{2}{\sqrt{-g(y)}}\frac{ \delta^2 \Gamma }{\delta g_{\rho \sigma}(y)\delta g_{\mu\nu}(x)} \Bigg |_{\kappa=0}.\label{ueue}
\end{align}
This corresponds to defining the perturbation variable by
\begin{align}
	g_{\mu\nu} = a^2 (\eta_{\mu\nu} + \kappa \tilde h_{\mu\nu}),
\end{align}
or, equivalently, to the field redefinition
\begin{align}
	h^{\mu\nu} = - \tilde h^{\mu\nu} + \kappa \tilde h^{\mu\alpha}\eta_{\alpha \beta}\tilde h^{\beta\nu}.
\end{align}
The stress tensors and self-energies are related by
\begin{align}
	a^6 \tilde T^{\mu\nu} &= a^2 \bar T^{\mu\nu},
\\
	a^{12}\tilde \Pi^{\mu\nu\rho\sigma} &=
	a^{4}\bar \Pi^{\mu\nu\rho\sigma}
	- a^2 (\bar T^{\mu\sigma} \eta^{\nu \rho} + \bar T^{\nu\sigma} \eta^{\mu \rho}+ \bar T^{\mu\rho} \eta^{\nu \sigma} + \bar T^{\nu\rho} \eta^{\mu \sigma}).
\end{align}
Note that indices are raised and lowered with $\eta^{\mu\nu}$ and $\eta_{\mu\nu}$.
Their natural positions are fixed by the covariant definitions, such as Eqs.~\eqref{unpi} and \eqref{ueue}. 
After some algebra, the Ward identities become
\begin{align}
	\partial_\nu (a^6 \tilde T^{\mu\nu}) - a^6 \eta_{\rho \sigma}\tilde T^{\rho\sigma} \partial^\mu \ln a = 0,
\end{align}
and
\begin{align}
	&\partial_\mu \circ (a^{12} \tilde \Pi^{\mu\nu\rho\sigma})
	+  \eta^{\rho \nu} \partial_\mu \circ( a^6 \tilde T^{\mu \sigma})
	+ \eta^{\sigma \nu} \partial_\mu \circ (a^6 \tilde T^{\mu \rho})
	-a^6 \tilde T^{\rho\sigma}\partial^\nu
	\notag\\
	&
	- 2 a^6 (\partial^\nu \ln a )\left( \tilde T^{\rho\sigma}
	+ \frac{1}{2} a^{6}\eta_{\alpha \beta}\tilde \Pi^{\alpha \beta \rho\sigma}\right)  = 0.\label{uepiid}
\end{align}
This operator identity acts on $\tilde h_{\rho\sigma}$.

\medskip
Another useful convention is
\begin{align}
	\check T^\mu{}_\nu &\equiv \left. \frac{2}{\sqrt{-g}}\frac{\delta \Gamma}{\delta g_{\mu \alpha}}g_{\alpha \nu}\right|_{\kappa = 0},\label{mixtau}
	\\
	\frac{a^4(x)}{a^4(y)}\left[\check \Pi^\mu{}_\nu{}^\rho{}_\sigma\delta^{(4)}(x-y)\right] &\equiv \left. \frac{2}{\sqrt{-g(x)}}\frac{2}{\sqrt{-g(y)}}   \frac{\delta}{\delta g_{\rho \beta}(y)}  \left[ \frac{\delta \Gamma}{\delta g_{\mu \alpha}(x)}g_{\alpha \nu}(x) \right]g_{\beta \sigma}(y)\right|_{\kappa = 0,{\rm sym.}}.\label{mixpi}
\end{align}
The subscript ``sym.'' implies that we symmetrize the variation in the arguments.
This convention corresponds to defining the perturbation variables through the exponential:
\begin{align}
	g^{\mu\nu} = a^{-2}(e^{-\kappa \check h})^\mu{}_\rho \eta^{\rho\nu},~g_{\mu\nu} = a^{2} \eta_{\mu \rho}(e^{\kappa \check h})^\rho{}_\nu,\label{defexpmet}
\end{align}
or the quadratic field redefinition
\begin{align}
	h^\mu{}_\nu = -\check h^\mu{}_{\nu} + \frac{\kappa}{2}\check h^\mu{}_\alpha \check h^\alpha{}_\nu + \mathcal O(\kappa^2).
\end{align}
For traceless $\check h^\mu{}_\nu$ the volume element is unperturbed, although matter remains coupled to the tensor perturbation through the metric.
The stress tensors and self-energies are related by
\begin{align}
	a^4 \check T^{\mu}{}_{\nu} &= a^2 \bar T^{\mu}{}_{\nu},
\\
	a^{8}\check \Pi^{\mu}{}_{\nu}{}^\rho{}_\sigma &= a^{4}\bar \Pi^{\mu}{}_{\nu}{}^\rho{}_\sigma - \frac{1}{2}a^2 (\bar T^{\mu}{}_{\sigma} \delta^{\rho}{}_{\nu} + \bar T_{\nu \sigma} \eta^{\mu \rho} + \bar T^{\mu \rho} \eta_{\nu \sigma} + \bar T^{\rho}{}_{\nu} \delta^{\mu}{}_{\sigma}).\label{defcheck}
\end{align}
Then, in this convention, we have the following identities:
\begin{align}
	\partial^\mu ( a^4 \check T_{\mu\nu} ) - a^4 \eta^{\rho\sigma}\check T_{\rho\sigma} \partial_\nu \ln a = 0,
\end{align}
and
\begin{align}
	0=&\partial_\mu \circ (a^8 \check \Pi^\mu{}_\nu{}^\rho{}_\sigma)
  -a^8 (\partial_\nu   \ln a )\check \Pi^\mu{}_{\mu}{}^\rho{}_\sigma    
  -a^4\check T^\rho{}_\sigma\partial_\nu
 \notag 
 \\
 &+ \frac{1}{2}  (\delta^{\rho}{}_{\nu} \partial_\mu \circ ( a^4 \check T^{\mu}{}_{\sigma} ) + \eta_{\nu \sigma } \partial_\mu \circ (a^4 \check T^{\mu \rho} ))
 \notag\\
 &- \frac{1}{2} (\partial^\rho \circ (a^4 \check T_{\nu \sigma} ) + \partial_\sigma \circ ( a^4 \check T^{\rho}{}_{\nu} )).\label{upunko}
\end{align}
This operator identity acts on $\check h^\sigma{}_\rho$. Neither identity assumes that $a^2\bar T_{\mu\nu}$ is constant.
These conventions are field redefinitions rather than gauge choices.
The split between the retarded and the contact response is convention dependent as well: earlier works define the perturbation through $\sqrt{-g}\,g^{\mu\nu}$~\cite{Frenkel:1990dw,Brandt:1991qn,Brandt:1993bj}, in which the minimally coupled kinetic term is exactly linear for a scalar field, so that it generates no four-point vertex and no contact response at leading HTL order.
We do not discuss this convention here.
After a field redefinition, the identities take different forms, although their physical content $\delta_\xi \Gamma=0$ is unchanged.

\medskip
This field-redefinition dependence of the self-energy has been recognized since Ref.~\cite{Rebhan:1990yr} and discussed extensively in Refs.~\cite{Brandt:1991qn,Brandt:1993bj}.
Our own pairing is Eq.~\eqref{pidefdiag} with Eq.~\eqref{mywi}.
For example, if one defines the self-energy via Eq.~\eqref{ueue}, then the appropriate identity is Eq.~\eqref{uepiid}.
By contrast, the convention~\eqref{mixpi} leads to the identity~\eqref{upunko}.
After matching conventions, Eq.~(4.3) of Ref.~\cite{Rebhan:1990yr} corresponds to Eq.~\eqref{uepiid} on a flat, homogeneous background.
Ref.~\cite{Brandt:1993bj} constructs a combination that is independent of the field parametrization within the class satisfying Eq.~(3.9) of that reference, through the trace subtraction
\begin{align}
\tilde \Pi^{\mu\nu\rho\sigma} \to \tilde \Pi^{\mu\nu\rho\sigma}
-\frac{1}{4}\eta_{\alpha\beta}\Big(&\eta^{\mu\rho}\tilde \Pi^{\alpha\beta\nu\sigma}
+\eta^{\mu\sigma}\tilde \Pi^{\alpha\beta\nu\rho}
\notag\\
&+\eta^{\nu\rho}\tilde \Pi^{\alpha\beta\mu\sigma}
+\eta^{\nu\sigma}\tilde \Pi^{\alpha\beta\mu\rho}\Big).
\end{align}
For conformal matter, the Weyl identity shows that this transformation agrees, at quadratic order, with the field redefinition from $\tilde h_{\mu\nu}$ to $\check h^\mu{}_\nu$, with the corresponding index positions and powers of $a$ taken into account.
The same trace subtraction for $\bar \Pi^{\mu\nu\rho\sigma}$ yields $\check \Pi^{\mu\nu\rho\sigma}$.
This corresponds to the field redefinition from $h^{\mu\nu}$ to $\check h^\mu{}_\nu$.
The transformed kernel must therefore be paired with the Ward identity for the exponential parametrization.

These constructions are all consistent with diffeomorphism invariance, but the resulting kernels cannot by themselves be identified with the physical self-energy governing the propagation of metric perturbations.
On an off-shell background, the quadratic kernel generally depends on the choice of field variables, whereas physical propagation cannot depend on this choice.
The appropriate procedure is to derive the perturbation equations from the full effective action, including the gravitational action, about a background satisfying its equations of motion.
The resulting physical dynamics is independent of the field parametrization.
We illustrate this procedure for gravitational waves in Sec.~\ref{seceffdyn}.

\section{Symmetry test}
\label{symtest}

In this section, we test our linear response construction~\eqref{deflineres} with the Ward identities and the Weyl identities of conformal matter.
Note that these identities constrain the response, but do not determine it.
On the other hand, local structures are strongly constrained.
We will show that a local state response is uniquely fixed at leading HTL order.
We then apply this result to the local Gibbs ensemble.
We further show that no local state response can be added to the collisionless response.

\subsection{Test of the collisionless response}\label{subsec:testcoll}
Given an equation of state and the ISO(3) symmetry of spacetime, one may solve the identity~\eqref{WI1} for $\bar T_{\mu\nu}$.
For the conformal scalar, $a^2 \bar T_{\mu\nu}$ is constant~\footnote{Note that the energy density is defined as $-T^0{}_0 \propto a^{-4}$.}, and the derivatives of $a^2 \bar T_{\mu\nu}$ can be eliminated from Eq.~\eqref{mywi}.

Next, from Eq.~\eqref{defpibar} one checks directly that $\bar T_{\mu\nu}$ and $\bar \Pi_{\mu\nu\rho\sigma}$ satisfy the Weyl identities~\eqref{wel1} and \eqref{wel2}. 
Direct contraction of the operator in Eq.~\eqref{defpibar} gives
\begin{align}
	\partial^{\mu}\circ(a^4 \bar \Pi_{\mu\nu\rho\sigma}) = a^2\left(\bar T_{\sigma\nu}\partial_\rho+\bar T_{\rho\nu}\partial_\sigma+\bar T_{\rho\sigma}\partial_\nu\right) ,\label{opeid}
\end{align}
which cancels the tadpole contribution in Eq.~\eqref{mywi}.
Any violation of Eqs.~\eqref{wel1} and~\eqref{wel2} feeds back into the diffeomorphism Ward identity~\eqref{mywi} through its last term, which we use below to constrain the state response.

\subsection{Local Gibbs ensemble}\label{subsec:localgibbs}
Let us now write the state response of the local Gibbs ensemble explicitly.
For the conformal scalar, we identify the local Gibbs ensemble by two conditions.
Its state response satisfies the Weyl identity.
It also satisfies the diffeomorphism Ward identity~\eqref{mywi} in the static limit, where time derivatives of the metric perturbation are negligible.
The latter condition implies that thermal equilibrium is defined in the static limit.

Let us start with the stress tensor in local thermal equilibrium
\begin{align}
	T^{\rm LG}_{\mu\nu}= (\rho+P)u_\mu u_\nu + P g_{\mu\nu}.
	\label{LGdef}
\end{align}
Generally $\rho$ and $P$ are perturbed by the metric.
Assuming that the total response is determined by the instantaneous local state, we identify $\bar\Pi^S_{\mu\nu\rho\sigma}$ through
\begin{align}
	T^{\rm LG}_{\mu\nu}=\left(1+\frac{\kappa}{2}h\right)\bar T_{\mu\nu}-\frac{\kappa}{2}a^2\bar\Pi^S_{\mu\nu\rho\sigma}h^{\rho\sigma}+\mathcal O(\kappa^2).
	\label{LGmatch}
\end{align}
We expand the local variables as
\begin{align}
	\rho = \bar\rho + \kappa\rho_{\alpha\beta}h^{\alpha\beta},\qquad
	u_\mu = -a\left(\delta^0_\mu + \kappa u_{\mu\alpha\beta}h^{\alpha\beta}\right).
	\label{LGexp}
\end{align}
Here $\rho_{\alpha\beta}$ and $u_{\mu\alpha\beta}$ are symmetric in $\alpha\beta$.
They may act on $h^{\alpha\beta}$ as operators.
The normalization condition $g^{\mu\nu}u_\mu u_\nu=-1$ gives
\begin{align}
	u_{0\alpha\beta} = \frac{1}{2}\delta^0_\alpha\delta^0_\beta.
	\label{LGnorm}
\end{align}
Expanding Eq.~\eqref{LGdef} to first order and comparing with Eq.~\eqref{LGmatch}, we find
\begin{align}
	a^2\bar\Pi^{S}_{\mu\nu\rho\sigma}
	=&-\frac{8}{3}a^2\rho_{\rho\sigma}\delta^0_\mu\delta^0_\nu
	-\frac{2}{3}a^2\rho_{\rho\sigma}\eta_{\mu\nu}
	+\frac{a^2\bar\rho}{3}\left(\eta_{\mu\rho}\eta_{\nu\sigma}+\eta_{\mu\sigma}\eta_{\nu\rho}\right)
	\notag\\
	&-\frac{8}{3}a^2\bar\rho\left(u_{\mu\rho\sigma}\delta^0_\nu+\delta^0_\mu u_{\nu\rho\sigma}\right)
	+\eta_{\rho\sigma}\bar T_{\mu\nu}.
	\label{PiSgen}
\end{align}
Note that $g^{\mu\nu}T^{\rm LG}_{\mu\nu}=0$ holds identically for $P=\rho/3$.
Then, with Eq.~\eqref{LGnorm}, the contraction over the first index pair gives the Weyl identity
\begin{align}
	\bar T_{\rho \sigma} - \frac{1}{2} a^2 \eta^{\alpha \beta}\bar \Pi^S_{\alpha \beta \rho\sigma} = 0
	\label{eqeq}
\end{align}
for any $\rho_{\alpha\beta}$ and $u_{i\alpha\beta}$.
However, Eq.~\eqref{PiSgen} is not symmetric under $\mu\nu\leftrightarrow\rho\sigma$ in general.
In Sec.~\ref{secdiff}, we required $a^4\bar\Pi^S_{\mu\nu\rho\sigma}$ to be symmetric.
Then the contraction over the second index pair must also give Eq.~\eqref{eqeq}, which requires
\begin{align}
	\eta^{\alpha\beta}\rho_{\alpha\beta} = 2\bar\rho,\qquad
	\eta^{\alpha\beta}u_{i\alpha\beta} = 0.
	\label{PiSweylcond}
\end{align}
These conditions do not fix $\rho_{\alpha\beta}$ and $u_{i\alpha\beta}$ completely.
The remaining freedom is fixed by the Ward identity.
At leading order in the HTL expansion, derivatives in $\rho_{\alpha\beta}$ and $u_{\mu\alpha\beta}$ appear only through powers of $\partial/T$, so we drop them.
Then $\rho_{\alpha\beta}$ and $u_{i\alpha\beta}$ are tensors built from $\eta_{\alpha\beta}$ and $\delta^0_\alpha$:
\begin{align}
	\rho_{\alpha\beta}=\bar\rho\left(A\,\delta^0_\alpha\delta^0_\beta+B\,\eta_{\alpha\beta}\right),\qquad
	u_{i\alpha\beta}=\frac{C}{2}\left(\delta^i_\alpha\delta^0_\beta+\delta^0_\alpha\delta^i_\beta\right),
	\label{LGansatz}
\end{align}
with dimensionless coefficients $A$, $B$, and $C$.
These coefficients may depend on time only through the background.
The scale factor can be removed by a Weyl rescaling of the conformal scalar.
The only scale in the matter sector is the temperature.
Hence a dimensionless coefficient can depend on time only through $\mathcal H/T$, which is subleading in the HTL expansion.
We therefore take $A$, $B$, and $C$ to be constant.
Then $a^4\rho_{\alpha\beta}$ and $u_{\mu\alpha\beta}$ are also constant, since $a^4\bar\rho$ is constant.
Both Eq.~\eqref{PiSweylcond} and the symmetry of Eq.~\eqref{PiSgen} reduce to $A=4B-2$.

Let us now impose the Ward identity on the local Gibbs response itself.
Namely, we take $T^{\rm LG}_{\mu\nu}$ as the full stress tensor and replace $\bar\Pi_{\mu\nu\rho\sigma}$ by $\bar\Pi^S_{\mu\nu\rho\sigma}$ in Eq.~\eqref{mywi}.
Since $a^2\bar T_{\mu\nu}$ and $a^4\bar\Pi^S_{\mu\nu\rho\sigma}$ are constant, the derivatives in Eq.~\eqref{mywi} act only on $h^{\rho\sigma}$.
The last term of Eq.~\eqref{mywi} vanishes by Eq.~\eqref{eqeq}.
The Ward identity then reduces to
\begin{align}
	a^4\bar\Pi^S_{\mu\nu\rho\sigma}\partial^\mu
	= a^2\left(\bar T_{\sigma\nu}\partial_\rho+\bar T_{\rho\nu}\partial_\sigma+\bar T_{\rho\sigma}\partial_\nu\right).
	\label{LGWI}
\end{align}
We substitute Eqs.~\eqref{LGnorm} and~\eqref{LGansatz} into both sides of Eq.~\eqref{LGWI}.
Comparing the components $(\nu,\rho,\sigma)=(i,j,j)$ with $j\neq i$, $(i,0,0)$, and $(0,i,0)$ for stationary perturbations with $\partial_0 h^{\rho\sigma}=0$, we obtain $B=0$, $A=-2$, and $C=-1$, respectively:
\begin{align}
	\rho_{\alpha\beta}=-2\bar\rho\,\delta^0_\alpha\delta^0_\beta,\qquad
	u_{i\alpha\beta}=-\frac{1}{2}\left(\delta^i_\alpha\delta^0_\beta+\delta^0_\alpha\delta^i_\beta\right).
	\label{LGTolman}
\end{align}
Substituting Eq.~\eqref{LGTolman} into Eq.~\eqref{PiSgen}, we uniquely determine the state response of the local Gibbs ensemble as
\begin{align}
	a^2\bar\Pi^{S}_{\mu\nu\rho\sigma}
	=&\ \frac{9}{2a^2\bar\rho}\,\bar T_{\mu\nu}\bar T_{\rho\sigma}
	-\frac12\left(\bar T_{\mu\nu}\eta_{\rho\sigma}+\eta_{\mu\nu}\bar T_{\rho\sigma}\right)
	\notag\\
	&+\eta_{\mu\rho}\bar T_{\nu\sigma}+\eta_{\mu\sigma}\bar T_{\nu\rho}+\eta_{\nu\rho}\bar T_{\mu\sigma}+\eta_{\nu\sigma}\bar T_{\mu\rho}
	\notag\\
	&+\frac{a^2\bar\rho}{6}\,\eta_{\mu\nu}\eta_{\rho\sigma}
	-\frac{a^2\bar\rho}{3}\left(\eta_{\mu\rho}\eta_{\nu\sigma}+\eta_{\mu\sigma}\eta_{\nu\rho}\right),
	\label{PiSsym}
\end{align}
where $a^2\bar\rho=\bar T_{00}$.

Note that Eq.~\eqref{PiSsym} satisfies the diffeomorphism Ward identity~\eqref{mywi} only in the static limit.
For $\partial_0 h^{\rho\sigma}\neq0$, the terms in which $\partial^\mu$ acts along $\delta^0_\mu$ are not balanced by the tadpole.
Explicitly, we find
\begin{align}
	\partial^\mu\circ\big(a^4\bar \Pi^S_{\mu\nu\rho\sigma}\big) &= a^2\left(\bar T_{\sigma\nu}\partial_\rho+\bar T_{\rho\nu}\partial_\sigma+\bar T_{\rho\sigma}\partial_\nu\right)
	\notag\\
	&\quad+\frac{4}{3}a^4\bar\rho\left(6\,\delta^0_\nu\delta^0_\rho\delta^0_\sigma+\delta^0_\nu\eta_{\rho\sigma}+\eta_{\nu\rho}\delta^0_\sigma+\eta_{\nu\sigma}\delta^0_\rho\right)\partial^0.
	\label{mywi2}
\end{align}
Hence the local Gibbs ensemble is consistent only with time-independent diffeomorphisms.
This is because we only consider the metric dependence of the stress tensor in Eq.~\eqref{LGexp}.
In the hydrodynamical setting, we introduce a hydrodynamic mode that nonlinearly realizes the diffeomorphism invariance, as we discuss in Sec.~\ref{subsec:hydro}.

\subsection{Origin of the superhorizon growth}\label{subsec:origingrowth}
Let us diagnose the origin of the growth in the superhorizon evolution found in our earlier treatment~\cite{Ota:2023iyh}.
The origin of its secular character is discussed in Sec.~\ref{seceffdyn}.
Our earlier prescription used the local Gibbs state response for TT perturbations.
Eq.~\eqref{PiSsym} extends it to arbitrary components of $h^{\rho\sigma}$.
This prescription was motivated by the interpretation that the linear response is computed around an instantaneous local equilibrium configuration since the free-streaming sector was in a local equilibrium state at some stage in the early universe.
We also argued there that the one-loop local contact diagram should be subtracted using the perturbed quantum state, since gravity can be locally eliminated by the equivalence principle.
In fact, we used the convention~\eqref{defexpmet} in the TT gauge, and the TT component of $\hat T^C_{\mu\nu}$ in that convention can be subtracted from the state response $\bar \Pi^S_{\mu\nu\rho\sigma}$
 in the local Gibbs state.
Then, the remaining response in that calculation was solely expressed by the genuine nonlocal response $\hat T^R_{\mu\nu}$, which we interpreted as the appropriate contribution.
However, this state is not compatible with the Ward identity~\eqref{mywi}.
Since $\bar\Pi_{\mu\nu\rho\sigma}$ already satisfies Eq.~\eqref{mywi} with the tadpole, the Ward identity for the total response $\bar \Pi_{\mu\nu\rho\sigma} +  \bar \Pi^S_{\mu\nu\rho\sigma}$ reduces to
\begin{align}
	\partial^\mu\circ(a^4\bar\Pi^S_{\mu\nu\rho\sigma})-2a^2(\partial_\nu\ln a)\bar T_{\rho\sigma}=0.
\end{align}
For $\nu=0$, Eq.~\eqref{mywi2} yields
\begin{align}
	\bar T_{0\sigma}\partial_\rho+\bar T_{0\rho}\partial_\sigma-\bar T_{\rho\sigma}\left(3\partial_0 + 2\mathcal H\right) = 0,
\end{align}
as an operator acting on $h^{\rho\sigma}$.
The first two terms act only on $h^{0\rho}$.
For the spatial components $(\rho,\sigma)=(i,j)$, they drop out, and the condition reduces to
\begin{align}
	\bar T_{ij}(3\partial_0+2\mathcal H)=0.
\end{align}
Since $\bar T_{ij}\neq0$ and $\mathcal H\neq0$ in the radiation era, this cannot hold even for stationary perturbations with $\partial_0h^{ij}=0$.

The violation has two origins.
The terms with $\partial_0$ reflect that the local Gibbs ensemble is consistent only with time-independent diffeomorphisms.
The remaining terms, including the one with $\mathcal H$, survive even for stationary perturbations.
In this case, Eq.~\eqref{PiSsym} alone satisfies the Ward identity together with the tadpole.
However, $\bar\Pi_{\mu\nu\rho\sigma}$ already satisfies the identity with the same tadpole.
Adding $\bar\Pi^S_{\mu\nu\rho\sigma}$ therefore counts the tadpole-related response twice.
In fact, the static limit of Eq.~\eqref{defpibar}, $\partial_0\to0$ with the retarded prescription, coincides with Eq.~\eqref{PiSsym}.

Note that the violation involves only the trace $\delta_{ij}h^{ij}$, because $\bar T_{ij}\propto\delta_{ij}$.
More generally, the violation acts only on $h^{0\mu}$, the trace, and the divergence $\partial_j h^{ij}$.
It therefore vanishes for transverse-traceless perturbations.
However, the Ward identity must hold for arbitrary $h^{\rho\sigma}$.
Restricting a kernel that violates it to the TT sector does not make it consistent.
Moreover, $\bar\Pi^S_{\mu\nu\rho\sigma}$ still contributes to the TT sector.
For TT perturbations, Eq.~\eqref{PiSsym} gives
\begin{align}
	a^2\bar\Pi^S_{ijkl}h^{kl}=\frac{2}{3}a^2\bar\rho\,h_{ij},
\end{align}
so that the state response adds
\begin{align}
	-\frac{1}{4}\kappa^2a^2\bar\Pi^S_{ijkl}h^{kl}=-2\mathcal H^2h_{ij},
\end{align}
to the tensor equation of motion.
Here we used the Friedmann equation.
This negative local term outweighs the positive mass term that the collisionless response yields in the infrared limit (Sec.~\ref{seceffdyn}).
It is responsible for the growth.

\subsection{No-go for local state responses}\label{subsec:nogo}
More generally, no local state response can be added to $\bar\Pi_{\mu\nu\rho\sigma}$.
The argument assumes only that the additional response is local, isotropic, and of leading HTL order.
Since $\bar\Pi_{\mu\nu\rho\sigma}$ satisfies Eq.~\eqref{mywi} with the tadpole, an additional $\bar\Pi^S_{\mu\nu\rho\sigma}$ with the same tadpole must satisfy the homogeneous identity
\begin{align}
	\partial^\mu\circ\left(a^4\bar\Pi^S_{\mu\nu\rho\sigma}\right)-a^4(\partial_\nu\ln a)\,\eta^{\alpha\beta}\bar\Pi^S_{\alpha\beta\rho\sigma}=0
	\label{homWI}
\end{align}
for arbitrary $h^{\rho\sigma}$.
At leading order in the HTL expansion, a local state response carries no derivatives.
As in Eq.~\eqref{LGansatz}, it is then a combination of the five isotropic structures built from $\eta_{\alpha\beta}$ and $\delta^0_\alpha$ with constant coefficients,
\begin{align}
	a^4\bar\Pi^S_{\mu\nu\rho\sigma}
	=&\ a^4\bar\rho\Big[c_1\,\delta^0_\mu\delta^0_\nu\delta^0_\rho\delta^0_\sigma
	+c_2\left(\delta^0_\mu\delta^0_\nu\eta_{\rho\sigma}+\eta_{\mu\nu}\delta^0_\rho\delta^0_\sigma\right)
	+c_3\,\eta_{\mu\nu}\eta_{\rho\sigma}
	+c_4\left(\eta_{\mu\rho}\eta_{\nu\sigma}+\eta_{\mu\sigma}\eta_{\nu\rho}\right)
	\notag\\
	&+c_5\left(\delta^0_\mu\delta^0_\rho\eta_{\nu\sigma}+\delta^0_\mu\delta^0_\sigma\eta_{\nu\rho}+\delta^0_\nu\delta^0_\rho\eta_{\mu\sigma}+\delta^0_\nu\delta^0_\sigma\eta_{\mu\rho}\right)\Big].
	\label{PiSloc}
\end{align}
Substituting Eq.~\eqref{PiSloc} into Eq.~\eqref{homWI}, we find
\begin{align}
	0=&\ c_1\,\delta^0_\nu\delta^0_\rho\delta^0_\sigma\partial^0
	+c_2\left(\delta^0_\nu\eta_{\rho\sigma}\partial^0+\delta^0_\rho\delta^0_\sigma\partial_\nu\right)
	+c_3\,\eta_{\rho\sigma}\partial_\nu
	+c_4\left(\eta_{\nu\sigma}\partial_\rho+\eta_{\nu\rho}\partial_\sigma\right)
	\notag\\
	&+c_5\left(\delta^0_\rho\eta_{\nu\sigma}\partial^0+\delta^0_\sigma\eta_{\nu\rho}\partial^0+\delta^0_\nu\delta^0_\rho\partial_\sigma+\delta^0_\nu\delta^0_\sigma\partial_\rho\right)
	\notag\\
	&-\mathcal H\,\delta^0_\nu\left[\left(-c_1+4c_2+4c_5\right)\delta^0_\rho\delta^0_\sigma+\left(-c_2+4c_3+2c_4\right)\eta_{\rho\sigma}\right],
	\label{PiSlocWI}
\end{align}
as an operator acting on $h^{\rho\sigma}$.
Let us consider stationary perturbations, for which only the spatial derivatives act.
The terms with spatial derivatives and the terms without derivatives must then vanish separately.
The components $(\nu,\rho,\sigma)=(i,j,j)$ and $(i,i,j)$ with $j\neq i$ give $c_3=0$ and $c_4=0$.
The components $(i,0,0)$ and $(0,i,0)$ then give $c_2=0$ and $c_5=0$.
Finally, the $\nu=\rho=\sigma=0$ component of the last line gives $c_1=0$.
Thus a nonzero local state response cannot be added to the collisionless response, even in the static limit.
Any admissible addition to the collisionless response must therefore be nonlocal in time.
The local Gibbs response~\eqref{PiSsym} is admissible only when it replaces $\bar\Pi_{\mu\nu\rho\sigma}$, and only in the static limit.
On the other hand, nonlocal kernels satisfying Eq.~\eqref{homWI} and the Weyl identity exist in abundance, as noted at the beginning of this section.
The Ward identity alone does not exclude them.
In this paper, the medium starts in global equilibrium.
The unperturbed reference state carries no dependence on $h^{\mu\nu}$, and the bulk part of $\bar\Pi^S_{\mu\nu\rho\sigma}$ vanishes.

\subsection{Relation to hydrodynamics}\label{subsec:hydro}
In a perturbed spacetime, neither the choice of time slicing nor that of the constant-energy hypersurface is unique.
Accordingly, a diffeomorphism embedding of a canonical ensemble into the background spacetime is generally given by
\begin{align}
	\langle \hat T^I_{\mu\nu}\rangle^{\pi} = e^{\pounds_\pi}\bar T_{\mu\nu},
	\label{116}
\end{align}
where $\pounds_\pi$ denotes the Lie derivative along $\pi^\mu$.
The question is whether this freedom must be counted as a contribution to $\bar \Pi^S_{\mu\nu\rho\sigma}$.
Within the present framework, it need not be counted.
The stress tensor is already a functional of $h^{\mu\nu}$ that satisfies the Ward identities.
Counting Eq.~\eqref{116} again would double count the same response.
The situation is different outside the collisionless HTL framework.
Suppose that local equilibrium is imposed instantaneously and $\bar\Pi^S_{\mu\nu\rho\sigma}$ is used beyond the static limit.
Then the time-dependent diffeomorphisms can be restored by regarding Eq.~\eqref{116} as a St\"uckelberg transformation.
Here $\pi^\mu$ is promoted to a field that shifts under the diffeomorphism generated by $\xi^\mu$.
The Ward identity then no longer holds identically in $h^{\rho\sigma}$.
Instead, it becomes the equation of motion for $\pi^\mu$, namely the linearized energy-momentum conservation of the fluid.
We solve it together with the Hamiltonian and momentum constraints of the Einstein equation.
Here $\pi^0$ and $\pi^i$ correspond to the density and velocity perturbations.
This is the standard hydrodynamic treatment of cosmological perturbations.
It describes a collision-dominated fluid, and it replaces the collisionless response $\bar\Pi_{\mu\nu\rho\sigma}$ rather than adding to it.
For TT perturbations, Eq.~\eqref{PiSsym} alone cancels the pressure term, and gravitational waves propagate freely.

\section{Effective dynamics}
\label{seceffdyn}

Having completed the symmetry analysis of the bulk kernel, we turn to the effective dynamics of gravitational perturbations.
We check consistency in a known example: primordial tensor perturbations in a radiation-dominated universe.
The preceding construction requires no gauge fixing.
We use the transverse-traceless (TT) sector to illustrate the dynamics and compare with earlier work.
From here on, we focus on the physical TT tensor modes satisfying $h^{0\mu}=0$, $\partial_i h^{ij}=0$, and $\delta_{ij}h^{ij}=0$.

The kernel $\bar \Pi_{ijkl}$ is the medium's stress-tensor response consistent with the Ward identities; it is not yet the self-energy of the dynamical tensor variable.
The quantity governing the propagation of $h_{ij}$ is obtained only after inserting $\bar \Pi_{ijkl}$ into the Einstein equation, eliminating terms proportional to the background equations, and projecting to TT~\footnote{For a recent discussion on the graviton self-energy in the on-shell background, see also Ref.~\cite{Liu:2024utl}.}.
This is the procedure for reducing the self-energy to the on-shell background.
Analyses that start in flat space and feed $\bar \Pi$ directly into a mode equation bypass the last step and can misidentify the dynamics.
They miss the cancellation between the pressure term and the fast part in the linear response.

While a Minkowski background is simpler than FLRW, a flat-background gravitational HTL dispersion analysis is intrinsically incomplete.
Neglecting the expansion places one in a regime where thermal curvature effects are negligible against the spatial gradients of the graviton mode.
The self-energy one aims to study is then suppressed.
In practice, HTL effects are relevant near the horizon scale rather than deep inside it.
Hence we use the full real-time analysis for the dynamics near the Hubble horizon, and use the adiabatic approximation only to exhibit the mode structure.

\subsection{Integro-differential equation for the tensor mode}
In the FLRW background, the Einstein tensor in Eq.~\eqref{einseq} is
\begin{align}
	G_{00} &= 3\mathcal H^2, \\
	G_{ij} &= - \frac{\kappa}{2} (h_{ij}'' + 2 \mathcal H h_{ij}' - \nabla^2 h_{ij}) 
	- \left( \mathcal H^2 + 2 \mathcal H' \right) (\delta_{ij} - \kappa h_{ij}).
\end{align}
We then solve Eq.~\eqref{einseq} order by order in cosmological perturbations.
The unperturbed stress tensor is in the perfect-fluid form~\eqref{defhomstten}.
The background equations of motion are
\begin{align}
	3\mathcal H^2 &= \frac{\kappa^2}{4}a^2 \rho,\label{fdeq1}
	\\
	 \mathcal H^2 + 2 \mathcal H' &=- \frac{\kappa^2}{4} a^2 P .\label{fdeq2}
\end{align}
The first-order equations reduce to
\begin{align}
	 h_{ij}'' + 2 \mathcal H h_{ij}' - \nabla^2 h_{ij} 
	 + \kappa^2 a^2 \left[\frac{P}{2}  h_{ij} - \frac{1}{4}\bar \Pi^{\rm TT}_{ijkl} h^{kl} \right] = 0. \label{defEMT344s}
\end{align}
Note that the factor $\kappa$ in the definition of $h^{\mu\nu}$ sets the gravitational coupling in Eq.~\eqref{def:Hint}, while the explicit powers of $\kappa$ in Eq.~\eqref{defEMT344s} do not by themselves specify the order in the perturbations because the Einstein equation introduces the additional gravitational coupling to the stress tensor.
This is obvious from Eqs.~\eqref{fdeq1} and \eqref{fdeq2}, as the background equations already carry $\kappa^2$.
Since the background is time dependent, we Fourier transform only in space, so a dispersion relation in the usual sense is not defined.
We write the metric perturbation as
\begin{align}
	h_{ij}(\tau,\vec x) = \sum_{s=\pm}\int \frac{d^3k}{(2\pi)^3}e^{i\vec k \cdot \vec x}e^{(s)}_{ij}(\vec k)h^{(s)}(\tau, \vec k).
\end{align}
We normalize the graviton circular polarization tensors as
\begin{align}
	e^{ij}_{(s)}(\vec k) e^{(s')}_{ij}(-\vec k) = \delta_{(s)}^{(s')}.
\end{align}
With this normalization, one finds 
\begin{align}
e^{ij}_{(s)}(\vec k) p_i p_j = \frac{p^2}{2} e^{2is\phi} (1-\mu^2),	
\end{align}
where $\mu$ is the cosine of the angle between $p_i$ and $k_i$, and $\phi$ is the azimuthal angle of $p_i$ around $k_i$.
The TT projection of the self-energy is then
\begin{align}
	\bar \Pi^{\rm TT}_{ijkl} h^{kl}= \bar \rho \int_{-1}^{1} \frac{d\mu}{2} \frac{\partial_0^2 + k^2}{(\partial_0 + i k \mu)^2} \frac{(1 - \mu^2)^2}{4} h_{ij}. \label{412s}
\end{align}
The kernel operator in Eq.~\eqref{412s} is second order, which makes it difficult to invert.
We find a useful formula that reduces this second-order operator to first order:
\begin{align}
	-\frac{3}{4}  \int_{-1}^{1} \frac{d\mu}{2}
	(1 - \mu^2)^2\frac{ \partial_0^2+ k^2}{(\partial_0 + i k \mu)^2} = -2 +3 \int_{-1}^{1} \frac{d\mu}{2}
	(1 - \mu^2)^2 \frac{\partial_0 }{\partial_0 + i k \mu}  \label{result:self}.
\end{align}
Finding the kernel function is therefore equivalent to solving
\begin{align}
	\partial_0 h_{ij} = (\partial_0 + ik\mu)S_{ij},
\end{align}
for $S_{ij}$.
The solution to this differential equation is the sum of a homogeneous solution and a particular solution:
\begin{align}
	S_{ij} = e^{-ik\mu\tau} \sigma_{ij} + \int^{\tau}_{\tau_0} d\bar \tau \, e^{ik\mu (\bar \tau -\tau ) } \partial_{\bar \tau} h_{ij}(\bar \tau),
\end{align}
where $\sigma_{ij}$ is an integration constant.
Here $\sigma_{ij}$ is TT and must be fixed by initial data. It is constant in time but in general depends on $k$ and $\mu$.
We take it to be $\mu$ and $k$ independent for simplicity.
Consequently, we may pull $\sigma_{ij}$ out of the $\mu$ integral in Eq.~\eqref{result:self}.
The additional scale and angular dependence is interpreted as nontrivial initial data for the hard sector since $\partial_0+ik\mu$ is the transport operator for the phase space distribution, and $\sigma_{ij}$ is the amplitude of the homogeneous solution it annihilates.
In Sec.~\ref{secdiff} we restricted $\Gamma^R$ to the bulk and left such initial data to be reintroduced when integrating the equation of motion; this $\sigma_{ij}$ is the boundary contribution we dropped from $\bar \Pi^S_{ijkl}$.

The final expression is
\begin{align}
	-\frac{1}{4}a^2 \kappa^2 \bar \Pi^{\rm TT}_{ijkl} h^{kl} &= \mathcal H^2 \Big(-2h_{ij} + 24K(k\tau) \sigma_{ij} + 24 \int_{\tau_0}^\tau d\bar \tau \, K(k(\tau - \bar \tau)) \partial_{\bar \tau} h_{ij} (\bar \tau) \Big), \label{56ss}
\end{align}
with
\begin{align}
	K(x) \equiv \frac{1}{16} \int_{-1}^{1} d\mu
	(1 - \mu^2)^2   e^{-ix \mu} = -\frac{\sin x}{x^3} - \frac{3 \cos x}{x^4} + \frac{3 \sin x}{x^5}.
\end{align}
The first term in Eq.~\eqref{56ss} cancels the pressure term in Eq.~\eqref{defEMT344s}.
The important thing is that the origin of this local term is the nonlocal response $\langle \hat T^R_{\mu\nu}\rangle$ since, for the conformal scalar with the convention~\eqref{mycon}, $\langle \hat T^C_{\mu\nu}\rangle$ vanishes for the TT modes.
We interpret this as the fast response of the hard sector reducing to a local term at leading HTL order.
The retarded condition then fixes $\sigma_{ij}$: at the initial time the memory integral in Eq.~\eqref{56ss} vanishes and, for $k\tau_0\ll1$, the nonlocal response reduces to $\frac{8}{5}\mathcal H^2\sigma_{ij}$, which must be absent because the medium has not yet responded, unless a free-streaming anisotropy is already present at $\tau_0$.
Therefore, we set $\sigma_{ij}=0$.
After straightforward algebra, one arrives at the effective linearized Einstein equation,
\begin{align}
 h_{ij}''+2\mathcal H\,h_{ij}'+k^2 h_{ij}&=-24\mathcal H^2 \int_{\tau_0}^{\tau}\! d\bar\tau\,K\!\big(k(\tau-\bar\tau)\big)\,\partial_{\bar\tau}h_{ij}(\bar\tau),\label{wbdm}
\end{align}
with spatial momentum $k=|\vec k|$ and the conformal time derivative $'$.
Then, the resulting integro-differential equation has the same form as that in Ref.~\cite{Weinberg:2003ur}.
The initial-condition ambiguity noted in Ref.~\cite{Rebhan:1994zw} corresponds here to the choice of $\bar \Pi^S_{ijkl}$ and of $\sigma_{ij}$.
As stated in Sec.~\ref{symtest}, the medium starts in global equilibrium.
The local bulk part of $\bar \Pi^S_{ijkl}$ then vanishes by Eq.~\eqref{homWI}.
The local Gibbs alternative is excluded by the Ward identity of Sec.~\ref{symtest}.
The retarded condition sets $\sigma_{ij}=0$.
Hence no free parameter remains.

\subsection{Real-time evolution}\label{subsec:realtime}
From the symmetry perspective, on superhorizon scales the adiabatic mode can be removed by a constant shift $h_{ij}\to h_{ij}+\epsilon_{ij}$ induced by a large diffeomorphism~\cite{Weinberg:2003sw}, and Eq.~\eqref{wbdm} respects this infrared symmetry.

\begin{figure}
	\centering\includegraphics[width=.48\linewidth]{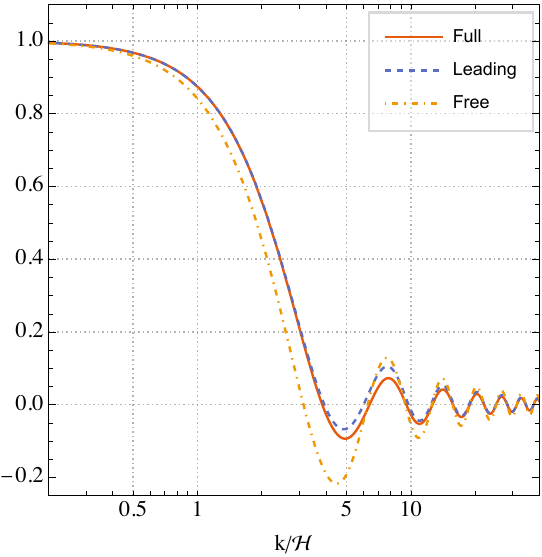}
	~
	\centering\includegraphics[width=.48\linewidth]{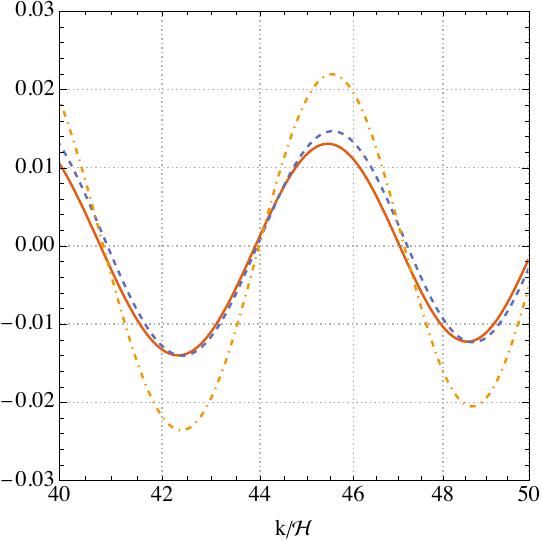}
	\caption{
Real-time evolution of the tensor-mode amplitude as a function of $k/\mathcal H$, obtained by directly integrating the integro-differential equation~\eqref{wbdm} (orange solid line), compared with the leading-order response in the gradient expansion~\eqref{wbdm3} (blue dashed line) and the free-graviton evolution described by the sinc function (yellow dot-dashed line).
In a radiation-dominated universe, the conformal time $\tau$ is related to the conformal Hubble parameter $\mathcal H$ by $\tau = \mathcal H^{-1}$.
All solutions are normalized to unity in the superhorizon regime, $k/\mathcal H\ll1$.
\textit{Left:} Immediately after horizon crossing, all curves closely track one another, with the linear-response solution being less suppressed than the free solution. This trend reverses near the first trough.
For the linear-response solutions, the near-horizon evolution is dominated by the leading-order response.
\textit{Right:} In the subhorizon regime, the envelopes approach a fixed ratio.
The full linear response exhibits additional damping, carried by the branch cut of the kernel.
}
\label{fig:realtime}
\end{figure}

To see the real-time evolution of $h_{ij}$, we solve Eq.~\eqref{wbdm} directly and compare it with the free solution, written in terms of the sinc function, in Fig.~\ref{fig:realtime}.
In the figure,  we normalize the amplitude to unity in the superhorizon regime where $k/\mathcal H\ll1$.
Immediately after horizon crossing, the linear-response solution~(orange solid line) is less damped than the free solution~(yellow dot-dashed line).
This trend reverses near the first trough, and the ratio of the two solutions is fixed in the deep subhorizon regime.
The leading HTL correction reduces the amplitude by about $40\%$ relative to the free solution.
As in Ref.~\cite{Weinberg:2003ur}, this suppression accumulates near horizon crossing.
Superhorizon modes are protected by the symmetry discussed above.
Taking the $k/\mathcal H \ll 1$ limit, one can localize the nonlocal integral in Eq.~\eqref{wbdm} and obtain
\begin{align}
 h_{ij}''+2\mathcal H\,h_{ij}'+ k^2 h_{ij} +\frac{8}{5}\mathcal H^2 (h_{ij} - h_{ij}(0)) =0.\label{wbdm3}
\end{align}
Here we take the limit $\tau_0\to0$ along the regular superhorizon solution, and $h_{ij}(0)$ denotes its initial value.
Thus, a time-dependent effective mass $\sqrt{8/5}\mathcal H$ arises, together with a source determined by the initial condition.
The leading-order gradient correction is also shown in Fig.~\ref{fig:realtime} by the blue dashed line.
The near-horizon behavior is dominated by Eq.~\eqref{wbdm3}.
The additional damping in the subhorizon regime arises from the next-to-leading-order nonlocal effect.
It is carried by the branch cut of the kernel.
We will return to this point in the next subsection.

Taken naively, the infrared limit reduces the retarded kernel as $\partial_0/(\partial_0+ik\mu)\to1$ and leaves only the mass term of Eq.~\eqref{wbdm3}, so that superhorizon modes are not conserved. 
This loss of memory accounts for the secular superhorizon evolution in our earlier work~\cite{Ota:2023iyh}.
The state response ruled out in Sec.~\ref{symtest} determines the sign of the local term and hence the growth.
The explicit retarded kernel removes the secular behavior.
The source in Eq.~\eqref{wbdm3} arises from its time boundary; it is not supplied by additional initial data for the distribution function.
The same kernel determines the evolution at finite $k$.

A related infrared analysis is given in Ref.~\cite{Glavan:2026wwl}, where the same positive plasmon mass is obtained~\footnote{Ref.~\cite{Glavan:2026wwl} appeared after the first version of this work.}.
Since they took the infrared limit for the two-point correlation, the secular issue remained.
The difficulty was addressed there by taking the hard particles to be in local equilibrium, perturbed at reheating by the long-wavelength tensor mode.
In the present language this corresponds to taking the infrared limit of the retarded kernel, discarding the history, and then restoring the time boundary as initial data of the distribution function, $\sigma_{ij}=-h_{ij}(0)$.
The physical reading behind it may well be correct.
What is restored, however, is the initial value alone, and the discarded history cannot be recovered afterwards.
The near-horizon evolution, where the medium effect is largest, is then out of reach.

\subsection{Adiabatic dispersion relation}

\begin{figure}
	\includegraphics[width=0.46\linewidth]{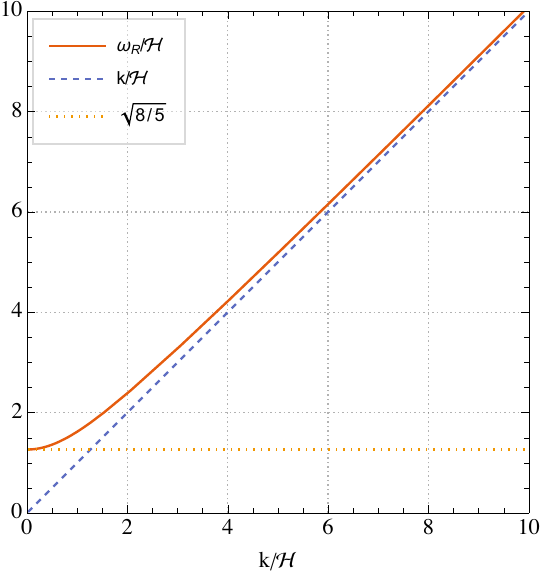}
	\includegraphics[width=0.49\linewidth]{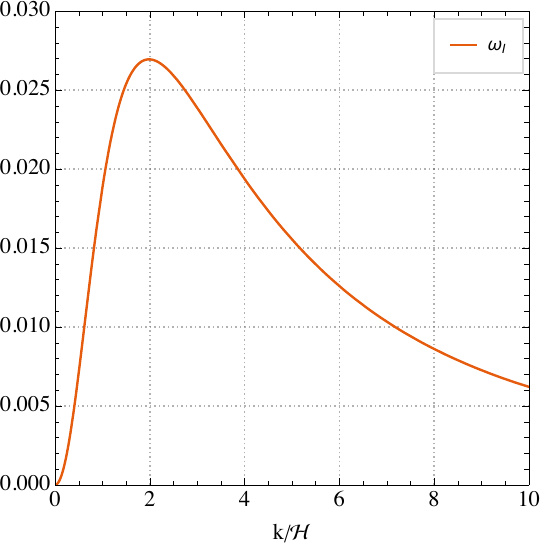}
	\caption{Numerical solution of the complex adiabatic dispersion relation~\eqref{wbdm4} for the complex frequency $\omega$ with $\mathcal H=1$.
\textit{Left:} The real part $\omega_R$ interpolates smoothly between the
massless dispersion $\omega=k$ at $k\gg\mathcal H$ and the plasmon mass $\omega\to\sqrt{8/5}\,\mathcal H$ at $k\ll\mathcal H$,
confirming Eqs.~\eqref{plasmonmass} and~\eqref{massnull}.
\textit{Right:} The imaginary part, $\omega_I$, approaches zero in both the $k\to\infty$ and $k\to0$ limits and exhibits a local extremum at $k\simeq2\mathcal H$.
We emphasize that this local feature of the \emph{adiabatic} dispersion relation does not by itself determine the actual evolution of cosmological perturbations in Fig.~\ref{fig:realtime} as the time boundary and time variation of $\mathcal H$ are crucial near the horizon scale.
}
	\label{fig1}
\end{figure}

The integro-differential equation~\eqref{wbdm} depends on cosmological initial data, hence on a specific time boundary.
However, one may wish to identify the mode structure of a graviton in a plane-wave expansion to see the collective property analogous to gauge theory at finite temperature.
Strictly, this is not possible in an expanding background, but one may consider an adiabatic approximation in which the background evolves slowly.
To separate out the Hubble friction due to the cosmic expansion, we normalize the tensor field as
\begin{align}
	v_{ij} = a h_{ij}.\label{normota}
\end{align}
Formally integrating Eq.~\eqref{412s} over $\mu$, we obtain
\begin{align}
 v_{ij}''+\left[ k^2 + \mathcal H^2  \mathcal I\left(\frac{iD_0}{k}\right) \right]v_{ij}=0,\label{wbdm2}
\end{align}
where we defined $D_0\equiv \partial_0 -\mathcal H$ and
\begin{align}
	 \mathcal I(y)&=
	 5y^2
-3y^4
+
\frac{3}{2}y(y^2-1)^2
\left[
\log(y+1)
-\log(y-1)
\right]
.\label{princ}
\end{align}
This kernel arises from the second term of Eq.~\eqref{result:self}, so that the fast response consistently cancels against the background pressure.
When we ignore the time dependence of $\mathcal H$, we find a ``dispersion relation''
\begin{align}
 \omega^2=k^2 + \mathcal H^2 \mathcal I\left(\frac{\omega -i\mathcal H}{k}\right).\label{wbdm4}
\end{align}
The solution of Eq.~\eqref{wbdm4} can be found numerically.
The result is shown in Fig.~\ref{fig1}.
In what follows, we discuss the features of the solution analytically.

\medskip
In gauge theories at finite temperature, a solution with $|{\rm Re}\,\omega| > k$ signals a positive plasmon mass shift; that is, massless waves in the zero-temperature limit acquire a gap in the dispersion relation in a finite-temperature plasma, indicating that the propagating mode is a collective excitation of the plasma.

Let us search for a self-consistent solution to Eq.~\eqref{wbdm4} by assuming $\omega \sim \mathcal H$.
Then, in the superhorizon limit, $k\ll \mathcal H$, we read the plasmon mass as
\begin{align}
	 \omega_{p}^2 =\lim_{k/\mathcal H\to 0} \mathcal H^2 \mathcal I\left(\frac{\omega-i\mathcal H}{k}\right)= \frac{8}{5} \mathcal H^2.\label{plasmonmass}
\end{align}
In the adiabatic description, the positive plasmon mass shift determines the minimum energy of the collective modes in the plasma.
If one naively applies plasma-physics intuition to the graviton case, a Hubble-scale plasmon mass would suggest that gravitons cannot exist on superhorizon scales because the frequency of a superhorizon mode when generated during inflation is below the gap.
On the other hand, the infrared limit of Eq.~\eqref{wbdm} is obtained as Eq.~\eqref{wbdm3}.
With cosmological boundary conditions the source term $\propto h_{ij}(0)$ in Eq.~\eqref{wbdm3} appears and keeps the superhorizon mode frozen before horizon reentry.

\medskip
Eq.~\eqref{result:self} has a pole at $\omega = k\mu$ and Eq.~\eqref{princ} has a branch cut.
This structure indicates a resonance between gravitons and plasma particles propagating in a specific relative direction.
After the rescaling~\eqref{normota}, the argument of $\mathcal I$ is $(\omega-i\mathcal H)/k$, so that the branch cut runs along $\mathrm{Im}\,\omega=\mathcal H$ with $|\mathrm{Re}\,\omega|\leq k$ and the branch points sit at $\omega=\pm k+i\mathcal H$: the cosmic expansion lifts the Landau cut off the real axis by $\mathcal H$.
Let us search for a self-consistent solution for $ \mathcal H < k\sim |\omega|$, meaning that the cosmic expansion is slower than the oscillation time scale of the collective mode.
We write $\omega = \omega_R + i \omega_I$ and search for a pole by solving Eq.~\eqref{wbdm4}.
We find a self-consistent solution in the subhorizon expansion, i.e., the expansion around a locally flat spacetime
\begin{align}
	\omega_\pm = \pm k  \pm \frac{\mathcal H^2}{k} + i\frac{\mathcal H^3}{k^2} +\left(\pm 3 \ln \left[\frac{\mathcal H}{2k}\right] \pm  5 - \frac{3\pi i}{2}\right) \frac{\mathcal H^4}{k^3} + \mathcal O\left(\frac{\mathcal H^5}{k^4}\right) .\label{520}
\end{align}
The two solutions are related by $\omega \to -\omega^*$, as required by the reality of $h_{ij}$, so that they share a common imaginary part.
When eliminating the Hubble expansion by Eq.~\eqref{normota}, the net contribution is a growing pole, as illustrated in Fig.~\ref{fig1}.
In fact, this anti-damping feature at $\mathcal O(\mathcal H^3/k^2)$ originates from the $-i\mathcal H$ shift in the argument of $\mathcal I$; it means that $h_{ij}$ decays slightly more slowly than $1/a$.
Note that we must place the time contour of the Laplace inverse transform above this growing pole to properly impose the retarded condition.

\begin{figure}
	\centering
	\includegraphics[width=0.7\linewidth]{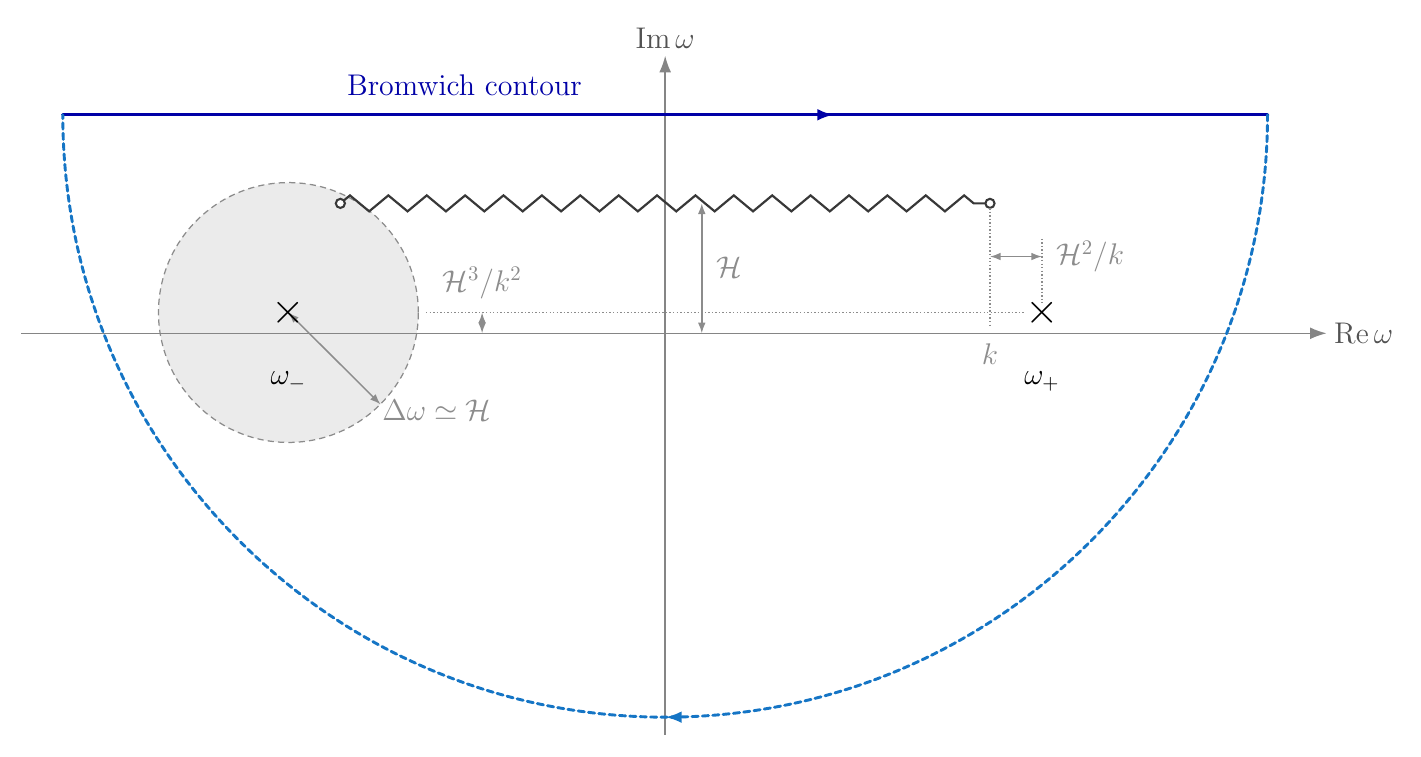}
	\caption{
Singularity and branch cut structure of the adiabatic response on the first Riemann sheet of the complex $\omega$ plane.
The cosmic expansion lifts the Landau cut to a height $\mathcal H$ above the real axis, with branch points at $\omega=\pm k+i\mathcal H$.
	The Bromwich contour lies above all singularities and is closed in the lower half plane.
The poles $\omega_\pm$ lie a distance $\mathcal H^{2}/k$ outside the branch points and $\mathcal H^{3}/k^{2}$ above the real axis.
The shaded disc of radius $\mathcal H$ around $\omega_-$ is the frequency resolution $\Delta\omega\sim\tau^{-1}\sim\mathcal H$ available within a Hubble time: the branch point falls inside it, so the pole and the cut are not separable within the adiabatic approximation.
All lengths are drawn to a common scale with $\mathcal H/k=2/5$.
	}
	\label{rdcnt}
\end{figure}

The imaginary part of the next-to-leading order correction, $\mathcal O(\mathcal H^4/k^3)$, carries a factor of $i\pi$ inherited from the logarithmic branch point at $\omega = \pm k + i\mathcal H$.
Since the mode has phase velocity larger than unity, $|{\rm Re}\,\omega|>k$, no massless particle keeps pace with the wave; its imaginary part then comes from the proximity of the branch point rather than from a genuine resonance.
The pole~\eqref{520} lies outside the cut and is reached from the upper half plane without crossing the cut, so that Eq.~\eqref{520} is the principal-branch pole, as illustrated in Fig.~\ref{rdcnt}.
Closing the Bromwich contour downward, we find that the retarded evolution receives contributions from the poles~\eqref{520} and from the branch cut, and neither of them alone fixes the amplitude.
In the real-time solution the suppression is accumulated around horizon entry and then saturates, which is the signature of the cut rather than of the pole: the imaginary part of Eq.~\eqref{520} would accumulate only $(\pi/2)(\mathcal H/k)^3$ in the adiabatic estimate over the range $k\tau\gg1$ where Eq.~\eqref{520} applies, far below the effect seen in Fig.~\ref{fig:realtime}.
In the subhorizon regime the poles lie only a distance $\mathcal H$ from the branch points, as illustrated in Fig.~\ref{rdcnt}.
Resolving them from the cut would then require $\tau \gg \mathcal H^{-1}$, which is beyond the reach of the adiabatic approximation.
The real-time equation~\eqref{wbdm} sums both contributions automatically, while Eq.~\eqref{520} locates the singularities and should not be read as a decay rate.
As $\mathcal H$ is lowered the pole merges with the branch point, and in the flat limit the collective mode disappears into the continuum.

The subhorizon expansion around the local Minkowski spacetime is valid for $\mathcal H/k \ll 1$.
The near-horizon regime is nonperturbative in this expansion, so we use the gradient expansion there.
As $k$ decreases, the two branch points $\omega=\pm k+i\mathcal H$ merge and annihilate, while the poles remain outside them and descend to the real axis at $\pm\omega_p$, as Fig.~\ref{fig1} shows. 
The infrared response is then carried by discrete poles alone, so the pole is unambiguously separated from the cut there, in contrast with the subhorizon regime.
Any finite order of the gradient expansion is local and therefore has no cut at all.
This is why the subhorizon series~\eqref{520} cannot be continued term by term to the infrared, and why the two expansions cannot be compared straightforwardly.

\medskip
Near $y=\pm1$, one obtains the asymptotic form
\begin{align}
	\mathcal I (\pm (1 - \epsilon)) = 2+2\epsilon+\mathcal O(\epsilon^2 \ln \epsilon),
\end{align}
which implies that 
\begin{align}
	\omega^2 = k^2 + 2\mathcal H^2 + \mathcal O(\mathcal H^3/k),\label{massnull}
\end{align}
for $k\gg \mathcal H$.
Gravitons in this limit acquire the subhorizon effective mass squared $2\mathcal H^2$.
Like the plasmon mass, this is not the mass in the Lagrangian.
Its effect relative to $k^2$ is suppressed for $k^2 \gg \mathcal H^2$.
The first real correction in Eq.~\eqref{520} can be interpreted as the mass correction in Eq.~\eqref{massnull}.
Eq.~\eqref{massnull} implies that the gravitational wave phase shifts by an amount of order $\mathcal H/k$ in every Hubble time $\mathcal H^{-1}$.
Eqs.~\eqref{plasmonmass} and~\eqref{massnull} are the values of the same kernel at the two ends of its argument, $\mathcal I(|y|\to\infty)=8/5$ and $\mathcal I(1)=2$.

\section{Conclusions}

In this manuscript, we examined how collective excitations of a thermal medium affect gravitational perturbations when the thermal soft scale coincides with the Hubble scale.
We first analyzed the response structure in quantum field theory and its symmetry constraints.
On this basis, we then examined how an adiabatic plasmon gap can coexist with conserved superhorizon tensor modes and what the dispersion relation tells us about their real-time cosmological evolution.

We performed the calculation in the canonical operator formalism without fixing a gauge, retaining both the nonlocal retarded response and the local contact response.
The contact response cancels the Schwinger terms produced by canonical time ordering.
We derived the diffeomorphism Ward identities for the bulk kernel in different field parametrizations, without imposing the background equations of motion.
Matching the identity to the convention in which the self-energy is defined is what makes the symmetry test meaningful.
This test rules out the state prescription responsible for the growth of the tensor power spectrum found in our earlier loop analyses~\cite{Ota:2023iyh,Ota:2024idm,Ota:2025yeu}.
Its secular character is removed by retaining the memory of the retarded kernel.
More generally, no local state response can be added to the collisionless kernel at leading HTL order.

We obtained the tensor-mode equation on a background that satisfies the Einstein equation.
The adiabatic dispersion relation identifies a gapped collective excitation, but its complex frequencies do not by themselves determine the cosmological tensor amplitude.
The amplitude follows from the real-time evolution equation, which reproduces Weinberg's damping in agreement with kinetic theory.
The adiabatic plasmon gap is compatible with conserved superhorizon tensor modes when the memory of the retarded response is retained.
For the initial state considered here, the symmetry constraints and the retarded boundary condition fix the remaining freedom in the response.

These results clarify both the information provided by the dispersion relation and the need for real-time evolution in a dynamical spacetime.
A natural extension is to study collective excitations beyond the hard-thermal-loop limit within the Schwinger-Keldysh framework.
The response of gravity to a thermal medium is relevant to cosmological observations~\cite{CMB-S4:2016ple,LISA:2017pwj,NANOGrav:2023gor,ET:2019dnz} and to the quantum dynamics of gravitational waves, including their interaction with the medium during inflation~\cite{Berera:1995ie,Berera:2008ar,Lyth:1995ka} and decoherence in the primordial plasma~\cite{Polarski:1995jg,Lombardo:2005iz,Burgess:2022nwu,deKruijf:2024ufs,Takeda:2025cye}.

\acknowledgments

The author thanks Yoshimasa Hidaka, Misao Sasaki, Toshifumi Noumi, Junyue Yang, Yi Wang, Sam Wong, and Takahiro Tanaka for useful discussions.
The author is grateful to Yuhang Zhu for their involvement in the initial phase of this work.
This work was supported in part by the National Natural Science Foundation of China under Grant Nos. 12403001 and 12547101, and New Chongqing YC Project CSTB2024YCJH-KYXM0083.	
The author is grateful to Yingying Lan for her generous support.

\bibliography{biblio.bib}{}
\bibliographystyle{unsrturl}

\end{document}